\documentclass[]{emulateapj}
\PassOptionsToPackage{hyphens}{url}\usepackage{hyperref}
\usepackage{natbib}
\usepackage{amssymb}
\usepackage{color}
\usepackage{amsmath,mathtools}
\usepackage{epsfig}
\usepackage[FIGTOPCAP]{subfigure}
\usepackage{afterpage}
\usepackage{enumerate}
\usepackage{multirow}
\usepackage{verbatim}
\usepackage{relsize}
\usepackage{tikz}
\usetikzlibrary{shapes.geometric, arrows}
\usepackage{morefloats}
\usepackage{wasysym}
\usepackage{rotating}

\usepackage{listings}
\definecolor{codegreen}{rgb}{0,0.6,0}
\definecolor{codegray}{rgb}{0.5,0.5,0.5}
\definecolor{codepurple}{rgb}{0.58,0,0.82}
\definecolor{backcolour}{rgb}{0.98,0.98,0.95}
\lstdefinestyle{mystyle}{
    commentstyle=\color{codegreen},
    keywordstyle=\color{magenta},
    numberstyle=\tiny\color{codegray},
    stringstyle=\color{codepurple},
    breakatwhitespace=false,         
    breaklines=true,                 
    captionpos=b,                    
    keepspaces=true,                 
    numbersep=5pt,                  
    showspaces=false,                
    showstringspaces=false,
    showtabs=false,                  
    tabsize=2
}
\newcommand{\noop}[1]{}

\newcommand{\Kp}{\ensuremath{Kp}}
\DeclareMathOperator*{\argmin}{arg\,min}

\shorttitle{EVEREST 2.0}
\shortauthors{Luger et al. 2016}

\begin{document}

\title{An update to the EVEREST K2 pipeline:\\ Short cadence, saturated stars, and Kepler-like photometry down to $\Kp = 15$}
\author{Rodrigo Luger\altaffilmark{1,2}, Ethan Kruse\altaffilmark{1}, Daniel Foreman-Mackey\altaffilmark{1,3},\\
Eric Agol\altaffilmark{1,2}, Nicholas Saunders\altaffilmark{1}}
\altaffiltext{1}{Astronomy Department, University of Washington, Box 351580, Seattle, WA 98195, USA; \href{mailto:rodluger@uw.edu}{rodluger@uw.edu}}
\altaffiltext{2}{Virtual Planetary Laboratory, Seattle, WA 98195, USA}
\altaffiltext{3}{Sagan Fellow}

\begin{abstract}
We present an update to the \texttt{EVEREST} $K2$ pipeline that addresses various limitations
in the previous version and improves the photometric precision of the de-trended light curves.
We develop a fast regularization scheme for third order pixel level decorrelation (PLD) and
adapt the algorithm to include the PLD vectors of neighboring stars to enhance the predictive
power of the model and minimize overfitting, particularly for faint stars. We also modify PLD to work for saturated
stars and improve its performance on extremely variable stars. On average, \texttt{EVEREST 2.0}
light curves have 10--20\% higher photometric precision than those in the previous version,
yielding the highest precision light curves at all $\Kp$ magnitudes of any publicly available $K2$
catalog. For most $K2$ campaigns, we recover the original \emph{Kepler} precision to at least
$\Kp = 14$, and to at least $\Kp = 15$ for campaigns 1, 5, and 6. We also de-trend all short
cadence targets observed by $K2$, obtaining even higher photometric precision for these stars.
All light curves for campaigns 0--8 are available online in the \texttt{EVEREST} catalog, which
will be continuously updated with future campaigns.
\texttt{EVEREST 2.0} is open source and is coded in a general framework that can be applied
to other photometric surveys, including \emph{Kepler} and the upcoming \emph{TESS} mission.\\
%\vspace*{0.7in}
\end{abstract}


\section{Introduction}
\label{sec:intro}
The failure of the second of four reaction wheels in 2013 brought the original
\emph{Kepler} mission to a premature conclusion, as the spacecraft could no longer
achieve the fine pointing precision required for the groundbreaking
transiting exoplanet and stellar variability science that the first four years of the
mission allowed. Since 2014 the spacecraft has been operating in a new mode, known
as $K2$, using the solar wind and periodic thruster firings to mitigate drift
\citep{Howell14}. Despite these measures, raw $K2$ photometry is significantly
poorer than that of the original \emph{Kepler} mission. In order to enable
continuing precision science with $K2$, numerous pipelines have been developed
\citep[e.g.,][]{VanderburgJohnson14,Armstrong15,Lund15,Crossfield15,ForemanMackey15,Huang15,Aigrain16},
many producing light curves with precision approaching that of \emph{Kepler} for
bright stars.

In \cite{Luger16} (henceforth Paper I), we developed the first version of our 
$K2$ pipeline, \texttt{EVEREST} (EPIC Variability Extraction and Removal for Exoplanet
Science Targets).
We employed a variant of pixel level decorrelation (PLD) based on the method of \cite{Deming15},
a data-driven approach that uses a star's own pixel-level light curve to remove
instrumental effects. We showed that \texttt{EVEREST} recovered the original \emph{Kepler} 
precision for stars brighter than $Kepler$-band magnitude $\Kp \approx 13$, yielding higher 
average precision than any publicly available $K2$ catalog for unsaturated stars.

In this paper, we present an update to our pipeline, which we refer to as \texttt{EVEREST 2.0}.
By combining PLD with spacecraft motion information obtained from nearby stars,
this update improves the precision of $K2$ light curves at all magnitudes relative to version \texttt{1.0} and addresses
certain issues with overfitting. \texttt{EVEREST 2.0} also reliably de-trends saturated
stars and stars observed in short cadence mode, obtaining comparable or even higher de-trending
power for these targets.

The paper is organized as follows: in \S\ref{sec:model} we derive the mathematical
framework of the \texttt{EVEREST 2.0} model, and in \S\ref{sec:implementation} we describe 
the implementation of our pipeline in detail. We present our results in \S\ref{sec:results}
and some additional remarks in \S\ref{sec:remarks}. In \S\ref{sec:using} we briefly discuss how
to use the \texttt{EVEREST} catalog and code, and in \S\ref{sec:conclusions} we 
summarize the work.

\section{The PLD Model}
\label{sec:model}
Here we describe the mathematical formulation of the \texttt{EVEREST}
pixel level decorrelation (PLD) model. In PLD, products of the fractional fluxes in each pixel
of the target aperture are used as regressors in a linear model:
\begin{align}
\label{eq:pldmodel}
\mathbf{m} = &\sum\limits_{i}                                 a_i     \frac{\mathbf{p}_{i}}                             { \sum\limits_{n}\mathbf{p}_{n}} +     \nonumber\\
             &\sum\limits_{i} \sum\limits_{j}                 b_{ij}  \frac{\mathbf{p}_{i}\mathbf{p}_{j}}               {(\sum\limits_{n}\mathbf{p}_{n})^2} +  \nonumber\\
             &\sum\limits_{i} \sum\limits_{j} \sum\limits_{k} c_{ijk} \frac{\mathbf{p}_{i}\mathbf{p}_{j}\mathbf{p}_{k}} {(\sum\limits_{n}\mathbf{p}_{n})^3}.
\end{align}
In the expression above, $\mathbf{m}$ is the instrumental model and $\mathbf{p}_{i}$ is the flux in
the $i^{th}$ pixel; both are vector quantities defined at an array of times $\mathbf{t}$.
Each term corresponds to a different PLD order (first, second, and third) resulting from 
a Taylor expansion of the instrumental signal.
The $a_i$, $b_{ij}$, and $c_{ijk}$ are the linear weights of the model, which we seek
to obtain below. For a detailed discussion of the theory behind PLD, see 
\cite{Deming15} and Paper I. Below we simply discuss its mathematical implementation.

\subsection{Regularized Regression (\texttt{rPLD})}
\label{sec:regreg}
Given a timeseries $\mathbf{y}$ with $N_{dat}$ data points, we wish to find the linear
combination of $N_{reg}$ regressors that best fits the instrumental component of $\mathbf{y}$.
Expressed in vector form, our linear model is thus
\begin{align}
\label{eq:xdotw}
\mathbf{m} = \mathbf{X} \cdot \mathbf{w},
\end{align}
where $\mathbf{X}$ is the ($N_{dat} \times N_{reg}$) design matrix constructed from the set
of regressors (the fractional pixel fluxes in Equation~\ref{eq:pldmodel}) and $\mathbf{w}$ is 
the ($N_{reg} \times 1$) vector of weights (the set
$\{a_i, b_{ij}, c_{ijk}\}$). If $\mathbf{w}$ is known, the de-trended light curve is simply
\begin{align}
\label{eq:detrended}
\mathbf{y}' = \mathbf{y} - \mathbf{m}.
\end{align}
In Paper I, we obtained $\mathbf{w}$ by maximizing the likelihood function
\begin{align}
\label{eq:like0}
\log\mathcal{L}_0 =  &-\frac{1}{2} \left(\mathbf{y} - \mathbf{X} \cdot \mathbf{w}\right)^\top
                     \cdot
                     \mathbf{K^{-1}}
                     \cdot
                     \left(\mathbf{y} - \mathbf{X} \cdot \mathbf{w}\right) \nonumber\\  
                     &-\frac{1}{2} \log\left|\mathbf{K}\right| 
                     -\frac{N_{dat}}{2}\log 2\pi,
\end{align}
where $\mathbf{K}$ is the ($N_{dat} \times N_{dat}$) covariance matrix of the data and
$\mathbf{y}$ is the ($N_{dat} \times 1$) simple aperture photometry (SAP) flux. Since the 
number of third order PLD regressors can
be quite large (on the order of several thousand for a typical star, which is larger than the number
of data points), the problem is ill-posed, meaning that a unique solution does not exist and 
maximizing $\log\mathcal{L}_0$ is likely to lead to overfitting. We thus constructed $\mathbf{X}$
from the (smaller) set of $N_{pc}$ principal components of the PLD regressors. We chose $N_{pc}$
by performing cross-validation, which aims to maximize the predictive power of the model while
minimizing overfitting.

However, while principal component analysis (PCA) yields a set of components that captures 
the most variance among the PLD vectors, there is no guarantee that the principal components
are the ideal regressors in the PLD problem. Dimensionality reduction techniques such as PCA
inevitably lead to information loss, and so it is worthwhile to consider alternative
regression methods to fully exploit the potential of PLD.

A common regression method for ill-posed problems is regularization, in which a prior is imposed
on the values of the weights $\mathbf{w}$. Since overfitting occurs when $\mathbf{w}$ becomes very
large, regularization recasts the problem by adding a penalty term to the likelihood that 
increases with increasing $|\mathbf{w}|$. While many forms of regularization exist, we focus on
L2 regularization, since it has an analytic solution. Recently, \cite{Wang16} used L2
regularization in a ``causal pixel model'' to de-trend light curves from the original \emph{Kepler}
mission. L2 regularization is mathematically equivalent to
placing a Gaussian prior on each of the weights $\mathbf{w}$, so that the posterior likelihood function becomes
\begin{align}
\label{eq:like}
\log\mathcal{L} =  \log\mathcal{L}_0
                   -\frac{1}{2}
                   \mathbf{w}^\top \cdot \mathbf{\Lambda}^{-1} \cdot \mathbf{w}
                   -\frac{1}{2} \log\left|\mathbf{\Lambda}\right|,
\end{align}
where $\mathbf{\Lambda}$ is the ($N_{reg} \times N_{reg}$) regularization matrix,
which we choose to be diagonal for simplicity and for computational efficiency:
\begin{align}
\label{eq:Lambda}
\Lambda_{m,n} = \lambda_{n}^2\delta_{mn}.
\end{align}
Each element $\lambda_n^2$ in $\mathbf{\Lambda}$ is the variance of the 
zero-mean Gaussian prior on the weight of the corresponding column of the design matrix, 
$\mathbf{X}_{*,n}$. In practice we find that if we choose the $\lambda_{n}$ correctly, this model has 
a higher predictive power than the PCA model adopted in Paper I.

Given this formulation, our task is to find the weights $\mathbf{\hat{w}}$ that maximize 
the posterior probability 
$\mathcal{L}$. Differentiating 
Equation~(\ref{eq:like}) with respect to $\mathbf{w}$, we get
\begin{align}
\label{eq:gradlike}
\frac{\mathrm{d}\mathbf{\log\mathcal{L}}}{\mathrm{d}\mathbf{w}} &= 
\mathbf{X}^\top \cdot \mathbf{K}^{-1} \cdot \mathbf{y} \nonumber\\
&- \left( \mathbf{X}^\top \cdot \mathbf{K}^{-1} \cdot \mathbf{X} + \mathbf{\Lambda}^{-1} \right) \cdot \mathbf{w}.
\end{align}
By setting this expression equal to zero, we obtain the maximum \emph{a posteriori} prediction 
for the weights,
\begin{align}
\label{eq:what}
\mathbf{\hat{w}} = 
\left( \mathbf{X}^\top \cdot \mathbf{K}^{-1} \cdot \mathbf{X} + \mathbf{\Lambda}^{-1} \right)^{-1}
\cdot \mathbf{X}^\top \cdot \mathbf{K}^{-1} \cdot \mathbf{y}
\end{align}
with corresponding model
\begin{align}
\label{eq:model_slow}
\mathbf{m} = 
\mathbf{X} \cdot
\left( \mathbf{X}^\top \cdot \mathbf{K}^{-1} \cdot \mathbf{X} + \mathbf{\Lambda}^{-1} \right)^{-1}
\cdot \mathbf{X}^\top \cdot \mathbf{K}^{-1} \cdot \mathbf{y}.
\end{align}
In what follows, we refer to the implementation of PLD with regularized regression as \texttt{rPLD}.

\subsection{Cross-validation}
\label{sec:crossval}
Similarly to Paper I, we solve for $\mathbf{\Lambda}$ by cross-validation. For each value
of $\mathbf{\Lambda}$, the model is trained on one part of the light curve (the training set)
and used to de-trend the other part of the light curve (the validation set); see
\S\ref{sec:impl_crossval} for details. The value of 
$\mathbf{\Lambda}$ that results in the minimum scatter in the validation set is then chosen for
the final de-trending step. 

In principle, each of the $\lambda_n$ in $\mathbf{\Lambda}$ could take on a different value, 
but solving for each one requires minimizing an $N_{reg}$-dimensional function and is not computationally tractable.
Instead, we simplify the problem by requiring that all regressors of the same order have the
same regularization parameter $\lambda$. Provided we write the third order design matrix in the form
\begin{align}
\label{eq:design}
\mathbf{X} = 
\left(
\begin{array}{ccc}
  \mathbf{X_1} & \mathbf{X_2} & \mathbf{X_3}
\end{array}
\right),
\end{align}
where $\mathbf{X_n}$ is the matrix of $n^\mathrm{th}$ order regressors, we may
express the regularization matrix as
\begin{align}
\label{eq:Lambda_block}
\mathbf{\Lambda} = 
\left(
\begin{array}{ccc}
  \mathbf{\Lambda_1}      &                       & \\
  &                       \mathbf{\Lambda_2}      & \\
  &                       &                       \mathbf{\Lambda_3} \\
\end{array}
\right)
\end{align}
where $\mathbf{\Lambda_n} = \lambda_{n}^2\mathbf{I}$ is the 
$n^\mathrm{th}$ order regularization matrix and $\lambda_{n}^2$ is the variance
of the prior on the $n^\mathrm{th}$ order regressors. 

A typical $K2$ star with 30 aperture pixels has $N_{reg} \sim\ $5,000 regressors and
$N_{dat} \sim\ $500 data points in each cross-validation light curve segment 
(see \S\ref{sec:impl_crossval}). Evaluating the matrix inverse in Equation~(\ref{eq:model_slow})
is thus computationally expensive, and becomes prohibitive during cross-validation,
since this must be done for every set of $\lambda_{n}$'s. Fortunately, we can reduce 
the number of calculations with some linear algebra. First, we apply the Woodbury matrix 
identity \citep[e.g.,][]{GolubVanLoan96} to Equation~(\ref{eq:model_slow}), obtaining
\begin{align}
\label{eq:model_woodbury}
\mathbf{m} =      \mathbf{X} \cdot \mathbf{\Lambda} \cdot \mathbf{X}^\top
                  \cdot
                  \left(
                  \mathbf{X} \cdot \mathbf{\Lambda} \cdot \mathbf{X}^\top + \mathbf{K}
                  \right)^{-1} 
                  \cdot
                  \mathbf{y}.
\end{align}
Next, we note that
\begin{align}
\label{eq:separable1}
\mathbf{X} \hspace{-2pt} \cdot \hspace{-2pt} \mathbf{\Lambda} \hspace{-2pt} \cdot \hspace{-2pt} \mathbf{X}^\top &= 
\left(
\begin{array}{ccc}
  \mathbf{X_1} & \mathbf{X_2} & \mathbf{X_3}
\end{array}
\right)
\left(
\begin{array}{ccc}
  \mathbf{\Lambda_1}\hspace*{-8pt}      &                                     & \\[2pt]
  &                                     \mathbf{\Lambda_2}\hspace*{-8pt}      & \\[2pt]
  &                                     &                                     \mathbf{\Lambda_3} \\
\end{array}
\right)
\left(
\begin{array}{c}
  \mathbf{X_1^\top} \\[2pt]
  \mathbf{X_2^\top} \\[2pt]
  \mathbf{X_3^\top} \\
\end{array}
\right) \nonumber\\[5pt]
&= \lambda_1^2 \mathbf{X_1} \hspace{-2pt} \cdot \hspace{-2pt} \mathbf{X^\top_1} +
   \lambda_2^2 \mathbf{X_2} \hspace{-2pt} \cdot \hspace{-2pt} \mathbf{X^\top_2} +
   \lambda_3^2 \mathbf{X_3} \hspace{-2pt} \cdot \hspace{-2pt} \mathbf{X^\top_3} \nonumber\\[5pt]
&= \sum_n \lambda_n^2 \mathbf{X^2_n},
\end{align}
where we have defined
\begin{align}
\label{eq:x2}
\mathbf{X^2_n} \equiv \mathbf{X_n} \cdot \mathbf{X^\top_n}.
\end{align}
We may thus re-write our maximum \emph{a posteriori} model as
\begin{align}
\label{eq:model}
\mathbf{m} = \sum_n \lambda_n^2 \mathbf{X^2_n}
                  \cdot
                  \left(
                  \sum_n \lambda_n^2 \mathbf{X^2_n} + \mathbf{K}
                  \right)^{-1} 
                  \cdot
                  \mathbf{y}.
\end{align}
The matrix that we must invert in Equation~(\ref{eq:model}) has dimensions ($N_{dat} \times N_{dat}$),
while that in Equation~(\ref{eq:model_slow}) is ($N_{reg} \times N_{reg}$). Since
$N_{reg} \sim 10N_{dat}$, casting the model in this form can greatly speed up the
computation. In practice, we pre-compute the three matrices $\mathbf{X^2_n}$ at the beginning
of the cross-validation step, so the only time-consuming operation in Equation~(\ref{eq:model})
is the inversion.

\subsection{Neighboring Stars (\texttt{nPLD})}
\label{sec:neighboring}
One of the downsides of PLD is that the regressors used in the linear model tend to be noisy.
This is particularly a problem for faint targets, whose PLD vectors are dominated by photon
noise. Their light is also distributed over fewer pixels, resulting in a smaller set of
vectors on which to regress. The effect of this is evident in Figure~10 of Paper I, which shows how
\texttt{EVEREST 1.0} light curves for faint ($\Kp \gtrsim 15$) stars are significantly noisier 
than those of the original \emph{Kepler} mission. This decrease in de-trending power at the
faint end affects most other $K2$ pipelines as well, as those usually regress on information
derived (either directly or indirectly) from the motion of the stellar image across the detector, which is similarly
noisy.

While the spacecraft motion (the dominant source of instrumental noise in $K2$) is imprinted at relatively 
low signal-to-noise ratio (SNR) on the light curves
of any one star, the collective light curves of all the stars on the detector encode
this information at very high SNR. Therefore, a straightforward way to improve the performance of PLD
for faint targets is to include this information in the design matrix. To this end, in \texttt{EVEREST 2.0} we
incorporate the PLD vectors of a set of other targets located on the same CCD module as the target of interest
when performing the regression.
We dub
this method \texttt{nPLD}, for \emph{n}eighboring
\emph{PLD}, and discuss its implementation in \S\ref{sec:impl_neighboring}. Our third-order design 
matrix (Equation~\ref{eq:design}) is now
\begin{align}
\label{eq:design_nPLD}
\mathbf{X} = 
\left(
\begin{array}{cccccc}
  \mathbf{X_1} & \mathbf{X_1'} & \mathbf{X_2} & \mathbf{X_2'} & \mathbf{X_3} & \mathbf{X_3'}
\end{array}
\right),
\end{align}
where $\mathbf{X_n'}$ is the design matrix constructed from the $n^{th}$ order PLD vectors of all the 
neighboring targets. For computational speed, we still solve for a single prior amplitude $\lambda_n$ 
for each PLD order, but in principle one could assign different priors to the neighboring vectors. We
discuss the implementation of \texttt{nPLD} in \S\ref{sec:implementation}.

\section{Implementation}
\label{sec:implementation}

\subsection{Light Curves}
\label{sec:impl_lightcurves}
As in Paper I, we downloaded all stars in the $K2$ EPIC catalog with long and/or short cadence
target pixel files and adopted aperture \#15
from the \texttt{K2SFF} catalog \citep{Vanderburg14,VanderburgJohnson14}. We masked all
cadences with \texttt{QUALITY} flags 1-9, 11-14, and 16-17, though we still compute the
model prediction on them. For campaigns 0-2, we remove the background signal as 
described in Paper I; for more recent campaigns, the background is removed by the \emph{Kepler}
team.

Next, we perform iterative sigma clipping to identify and mask outliers at 5$\sigma$.
During each iteration, we compute the linear (unregularized) PLD model and smooth it with
a Savitsky-Golay filter \citep{SavitskyGolay64}, then identify outliers based on a 
median absolute deviation (MAD) cut. We implement this outlier-clipping step at the
beginning of each cross-validation step (\S\ref{sec:impl_crossval}), each time computing
the model with a higher (regularized) PLD order, to progressively refine the outlier mask.

\subsection{GP Optimization}
\label{sec:impl_gp}
In order to compute the covariance matrix $\mathbf{K}$ for each target, we use a Gaussian
process (GP), as we did in Paper I. GP optimization can be costly, especially when
performing model selection over a range of possible kernels and optimizing many
hyperparameters simultaneously. For this reason, in Paper I we cut corners and performed
kernel selection based on fits to the autocorrelation function of the light curve, which
we also used to fix the timescale and/or period of those kernel(s). We then ran a nonlinear
minimizer to optimize the overall amplitude of the GP. In practice, this worked reasonably
well, but often failed for light curves dominated by high frequency stellar variability.
After much experimentation, we decided to forego the kernel selection step in favor of
using a single carefully optimized Mat\'ern-3/2 kernel with an added white noise term:
\begin{align}
\mathbf{K}_{ij} = \alpha \left(1 + \frac{\sqrt{3(t_i - t_j)^2}}{\tau}\right) e^{-\frac{\sqrt{3(t_i - t_j)^2}}{\tau}} + \sigma^2\delta_{ij},
\end{align}
where the hyperparameters $\sigma$, $\alpha$, and $\tau$ are the white noise amplitude,
red noise amplitude, and red noise timescale, respectively, and $t_i$ and $t_j$ correspond
to the timestamps of cadences $i$ and $j$. We initialize the hyperparameters at random
values and run a nonlinear optimizer to solve for the maximum likelihood (Equation~\ref{eq:like}),
keeping the PLD model parameters fixed; we repeat this process several times and retain the 
highest likelihood solution. As with outlier clipping, we progressively optimize the GP
at each of the three cross-validation steps, so that each time we train the GP on a 
light curve that is increasingly dominated by stellar variability (as opposed to instrumental
systematics).

In principle, the quasi-periodic kernels used in Paper I should be better suited to 
handling variable stars, but in practice we find that a properly optimized Mat\'ern-3/2 
kernel is flexible enough to fully capture the variability and prevent PLD overfitting.
We discuss this further in \S\ref{sec:remarks}.

\subsection{Breakpoints}
\label{sec:impl_breakpoints}
Because the instrumental noise properties are quite variable over the course of $K2$ 
campaigns, we find a significant improvement in the de-trending power of our regularized
regression model when we subdivide light curves into two or three segments. This is in contrast to
the PCA approach in Paper I, where we did not find it necessary to split the timeseries.
For all campaigns except 4 and 7, we add a single breakpoint in the light curve near the
mid-campaign point, where the spacecraft roll is at a minimum. For campaigns 4 and 7,
we find it necessary to insert two breakpoints. We cross-validate and de-trend each 
light curve segment separately and mend them at the end. In order to mitigate flux
discontinuities at the breakpoints, we train the model in each segment on an additional
100 cadences past the breakpoint to remove potential edge effects and offset the models
in each segment so that they align at the breakpoint.
While this method removes flux discontinuities, it can introduce discontinuities in the
derivative of the flux, showing up as spurious ``kinks'' in the light curve. We 
remove these in a post-processing step (\S\ref{sec:cbvs}).

\subsection{Neighboring Stars}
\label{sec:impl_neighboring}
In principle, the larger the number of neighboring PLD vectors we include in the
\texttt{nPLD} design matrix, the higher the de-trending power of our model. However, adding 
regressors significantly increases computing time, so we would like to instead select
a small set of high SNR regressors that capture most of the spacecraft
motion information. Moreover, since we employ a single prior for all $n^{th}$ order
regressors, adding many foreign PLD vectors effectively dilutes the contribution of
the target's own PLD vectors, which are the only ones that can correct instrumental
signals arising from local pixel sensitivity variations, and in practice results
in poorer quality light curves. After much experimenting, we obtain the highest average 
de-trending power when the number of neighboring stars is about ten. We therefore de-trend
each $K2$ target with the aid of the PLD vectors of ten randomly selected bright 
($11 \leq \Kp \leq 13$) stars on the same detector module as the target. To minimize
contamination of the target by outliers in its neighbors' fluxes, we linearly
interpolate over all neighbor data with flagged \texttt{QUALITY} bits. Finally, for
computational reasons, we neglect all cross terms of the form $\prod_{i \neq j} p_i p_j$,
where $p_i$ is the flux in the $i^{th}$ pixel,
when computing the neighbors' PLD vectors. Cross terms typically encode information
specific to the sets of pixels from which they are computed and aid in correcting
features such as thruster firing discontinuities \citep{Luger16}. Cross terms from
stars other than the target in question are therefore of little help in the de-trending
and can be safely neglected.

One potential pitfall of \texttt{nPLD} is that if the PLD assumptions break down for
any of the neighboring targets, the PLD regressors may become contaminated with
astrophysical information from that neighbor. This is not in general an issue, since
overfitting would only occur if an astrophysical signal in the target star and in
its neighbor had the same period and the same phase. However, in the (unlikely) case
that PLD fails for the neighboring star and this star happens to be an eclipsing binary 
or a transiting exoplanet host, it is possible that its transit signals could get imprinted onto
the target star's de-trended light curve, resulting in potential false positive planet
detections down the line. The two cases relevant to $K2$ in which PLD could fail in such
a way are for saturated stars and stars with bright contaminant sources in their
apertures \citep{Luger16}. As we show in \S\ref{sec:impl_saturated} below, it is 
straightforward to adapt PLD to work reliably for saturated stars, thereby circumventing
this issue. But while 
\texttt{EVEREST 2.0} is more robust against overfitting of crowded stars 
(\S\ref{sec:remarks}), highly crowded apertures remain an issue for PLD. When de-trending
with \texttt{nPLD}, we therefore select neighboring stars with no other known sources
in their apertures that are bright enough ($\Delta \Kp < 5$) to contaminate the PLD
vectors.

\subsection{Saturated Stars}
\label{sec:impl_saturated}
As discussed in Paper I, PLD typically fails for stars with saturated pixels, resulting
in overfitted light curves with artificially low scatter and suppressed astrophysical 
information (such as transits with significantly shallower depths). This happens because
saturated pixels contain nearly no astrophysical information, as the signal
overflows into adjacent pixels in the same column and is ultimately dumped into the
pixels at the top and bottom of the bleed trails; these ``tail'' pixels ultimately contain more 
astrophysical information than the other pixels in the aperture. Since PLD implicitly
assumes that astrophysical information is constant across the aperture, the method
breaks down for these stars, and PLD vectors from pixels in the saturated columns become capable
of fitting out the astrophysical information in the rest of the aperture.

An obvious workaround is to simply discard pixels in saturated columns from the set of PLD 
regressors. However, this does not work well in practice, since the remaining regressors
often have much lower SNR than the SAP flux and thus have low de-trending power.
We instead suggested in Paper I that collapsing saturated columns into single pixels --- by co-adding
the fluxes in each of the pixels and treating the resulting timeseries as a single PLD pixel ---
could reduce the effect of saturation, since charge is conserved along the bleed trail. While
this ensures PLD does not overfit, it leads to the loss of some of the information about the 
vertical motion of the stellar point spread function (PSF) across the detector. This leads to significantly poorer
de-trending, and therefore we did not employ this method in the first version of the pipeline.
However, we find that including the PLD vectors of neighboring stars in the design matrix 
(i.e., \texttt{nPLD}) effectively restores the information lost when saturated columns are 
collapsed, leading to high quality de-trended light curves of saturated stars.

In practice, we collapse all columns containing one or more pixels whose flux 
comes within 10\% of (or exceeds)
the pixel well depth for the corresponding detector channel during more than 2.5\% of the
timeseries. We obtained the well depths from
Table 13 of the Kepler Instrument Handbook.
\footnote{\url{archive.stsci.edu/kepler/manuals/KSCI-19033-001.pdf}} To avoid aperture
losses, we found it necessary to use aperture \texttt{K2SFF} \#19 (the largest of the PSF-based
apertures) for these stars. We further padded these apertures by two pixels at the top
and bottom of saturated columns to ensure that none of the target flux bled out of the
aperture; see \S\ref{sec:losses}.

\begin{figure}[hbt]
  \begin{center}
      \includegraphics[width=0.45\textwidth]{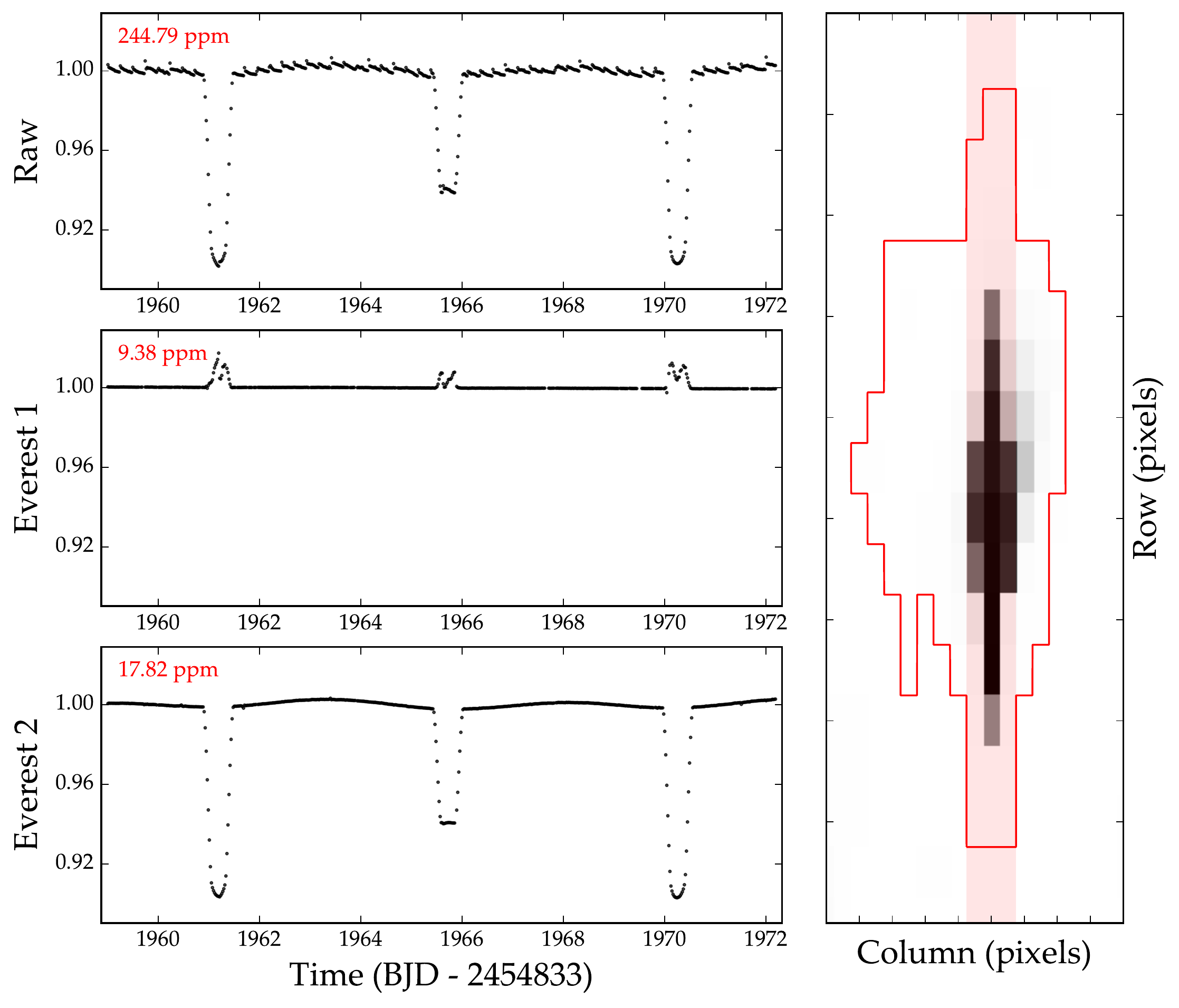}
       \caption{EPIC 202063160, a saturated \Kp = 9.2 campaign 0 eclipsing
       binary. Shown is a portion of the raw light curve (top), the light curve ``de-trended''
       with \texttt{EVEREST 1.0} (center), and the light curve de-trended with \texttt{EVEREST 2.0}
       (bottom); the $y$ axis in each of these plots is the normalized flux.
       The pixel image is shown at the right on a linear scale, with the adopted aperture
       contour indicated in red. The three columns highlighted in red contain saturated
       pixels. Despite a great improvement in the precision, \texttt{EVEREST 1.0} leads to severe
       overfitting, causing the eclipses to all but disappear. By collapsing saturated columns,
       \texttt{EVEREST 2.0} correctly de-trends saturated stars without overfitting.}
     \label{fig:saturated_star}
  \end{center}
\end{figure}

As an example, in Figure~\ref{fig:saturated_star} we plot the light curve of EPIC 202063160, a saturated
campaign 0 eclipsing binary. The raw light curve is shown at the top and the light curve
de-trended with \texttt{EVEREST 1.0} is shown at the center. Because three of the columns
in the aperture contain saturated pixels (right panel), \texttt{EVEREST 1.0} almost 
completely fits out the eclipses. With column-collapsed \texttt{nPLD} (bottom), the
eclipse is preserved and the instrumental signal is effectively removed.

\subsection{Short Cadence}
\label{sec:impl_shortcad}
We treat short cadence targets in much the same way as long cadence targets, with the exception
that we find it necessary to introduce more breakpoints in the light curves. This is
due primarily to computational reasons (short cadence $K2$ light curves are over $10^5$ cadences
in length; computing Equation~(\ref{eq:model}) for the entire light curve is not feasible).
Moreover, we find that noise on sub-30 minute timescales is only properly removed when
the size of the light curve segments is kept small. In practice, we find that on the order of
30 breakpoints result in the best de-trending. This might raise concerns of overfitting, but
since short cadence light curves contains 30 times more data than long cadence light curves,
and we split the latter into two segments, each of the short cadence segments has about twice
as many cadences as the long cadence ones.

The major downside of such a large number of segments are the discontinuities that could
be introduced at each breakpoint. As before, we overcompute the model into adjacent
segments and match the models at the breakpoints, but some de-trended light curves display
occasional jumps in either the flux or its derivative.

A second issue with short cadence light curves concerns deep transits and eclipses. As
we discussed in Paper I, PLD may attempt to fit out these features if they are not
properly masked, since doing so can result in a very large (but spurious) improvement in the
photometric precision.
With long cadence light curves, transit masking can be done by the user by simply re-computing
the model with the appropriate cadences masked, since the transits are sparse and their
presence does not significantly affect the cross-validation step. Moreover, outlier clipping
usually masks most deep transits anyways, so this is hardly ever a problem.
However, that is not the case with short cadence light curves, where transits and eclipses 
span upwards of fifty contiguous cadences. Since these features are so smooth, they are
not flagged as outliers. And since the transit signal is no longer sparse --- as it makes
up a substantial fraction of the light curve segment --- it is far more likely to bias
the cross-validation step. In practice, we find that this leads to substantial 
\emph{under}fitting of short cadence light curves with deep transits. As $\lambda_n$
increases, PLD begins to fit out the transit and the scatter in the validation set
grows, forcing the algorithm to select very low values of $\hat{\lambda}_n$ and resulting
in de-trended light curves that still contain significant instrumental signals.

We therefore explicitly mask all deep transits and eclipses in the short cadence light
curves \emph{before} the cross-validation step. Since only deep transits are likely to
bias the cross-validation, and since the number of short cadence light curves in each
campaign is relatively small, these can easily be identified by inspection.

\subsection{Cross-validation}
\label{sec:impl_crossval}
The principal step in the de-trending process is determining the prior amplitudes $\lambda_n$ in
Equation~(\ref{eq:model}), which we do by cross-validation. Our method is analogous to
that of Paper I, where we performed cross-validation to obtain the optimal number of principal
components to regress on. However, here we seek to optimize a three-dimensional function
$\sigma_\mathrm{v}(\lambda_1, \lambda_2, \lambda_3)$, where $\sigma_\mathrm{v}$ is the scatter 
in the validation set; this is a far more expensive calculation to do.
While we could employ a nonlinear optimization
algorithm (see below), in the interest of computational speed, we perform a simplification. Since we
expect the first order PLD regressors to contain most of the de-trending information,
with each successive PLD order providing a small correction term to the fit, we break
down the minimization problem into three separate one-dimensional problems.
First, we perform cross-validation on the first order PLD model by setting $\lambda_2 = \lambda_3 = 0$
to obtain the value of $\lambda_1$ that minimizes the validation scatter, $\hat{\lambda}_1$. We do this by computing the model
for each value of $\lambda_1$ in a logarithmically-spaced grid with 36 points in the range $[10^{0}, 10^{18}]$,
plus $\lambda_1 = 0$, and select the minimum (details below).
We then repeat this process on the second order model
by fixing $\lambda_1$ at this estimate and keeping $\lambda_3 = 0$. Finally, we solve
for $\hat{\lambda}_3$ by fixing the first and second order parameters at their optimum
values:
\begin{align}
\hat{\lambda}_1 &= \argmin \sigma_\mathrm{v}(\lambda_1)\bigg|_{\lambda_2 = 0,\               \lambda_3 = 0} \nonumber\\
\hat{\lambda}_2 &= \argmin \sigma_\mathrm{v}(\lambda_2)\bigg|_{\lambda_1 = \hat{\lambda}_1,\ \lambda_3 = 0} \nonumber\\
\hat{\lambda}_3 &= \argmin \sigma_\mathrm{v}(\lambda_3)\bigg|_{\lambda_1 = \hat{\lambda}_1,\ \lambda_2 = \hat{\lambda}_2}
\end{align}
It is important to note that there is no \emph{a priori} reason
that this method should yield the global minimum of $\sigma_\mathrm{v}$; in fact, it very
likely does not. However, we explicitly allow for $\lambda_n = 0$ in our grid search, and thus 
this approximation cannot lead to overfitting, as it will always prefer a lower-order PLD model to one
with higher scatter in the validation set. 

\begin{figure}[hbt]
  \begin{center}
      \includegraphics[width=0.47\textwidth]{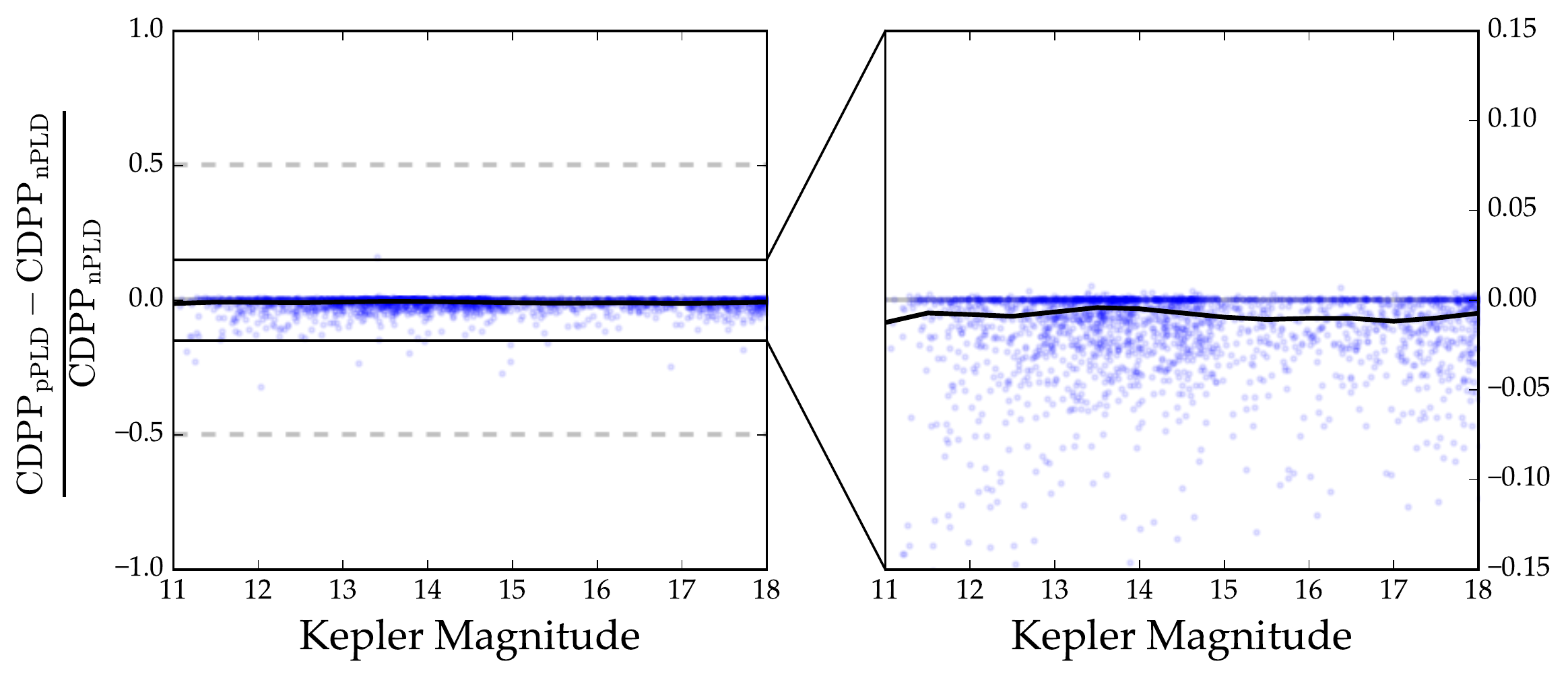}
       \caption{6 hr CDPP comparison between de-trending with \texttt{pPLD} and
                de-trending with \texttt{nPLD} for a sample of 2,700 randomly selected campaign 6
                stars. Plotted is the star-by-star difference in the CDPP values for each method,
                normalized to the \texttt{nPLD} CDPP (blue dots); stars with negative values have lower CDPP 
                when de-trended with \texttt{pPLD}. The black line is the median CDPP difference
                in 0.5 magnitude-wide bins. \texttt{pPLD} leads to an average improvement
                in the CDPP of ${\lesssim}1\%$.}
     \label{fig:pPLD}
  \end{center}
\end{figure}

As a proof of concept, we de-trended a sample of 2,700 randomly selected campaign 6 targets
with \texttt{nPLD} using this approximation in the cross-validation step. 
We then repeated the de-trending by
solving for the $\hat{\lambda}_n$ using Powell's method, initializing the solver at different
points in the vicinity of the \texttt{nPLD} solution and keeping the solution with the lowest
average 6 hr CDPP \citep[combined differential photometric precision;][]{Christiansen12} for each target; we dub this method 
\texttt{pPLD}. In Figure~\ref{fig:pPLD} we plot
the star-by-star CDPP difference between the two models, 
$\mathrm{(CDPP_{pPLD} - CDPP_{nPLD})/CDPP_{nPLD}}$. While for some stars the CDPP improves
substantially with \texttt{pPLD}, cross-validating with Powell's method 
leads to a less than one percent improvement in the CDPP on average. Given that this method
is more computationally expensive, we adopt the grid search method outlined above when
producing the \texttt{EVEREST 2.0} catalog.

\begin{figure}[hbt]
  \begin{center}
      \includegraphics[width=0.3\textwidth]{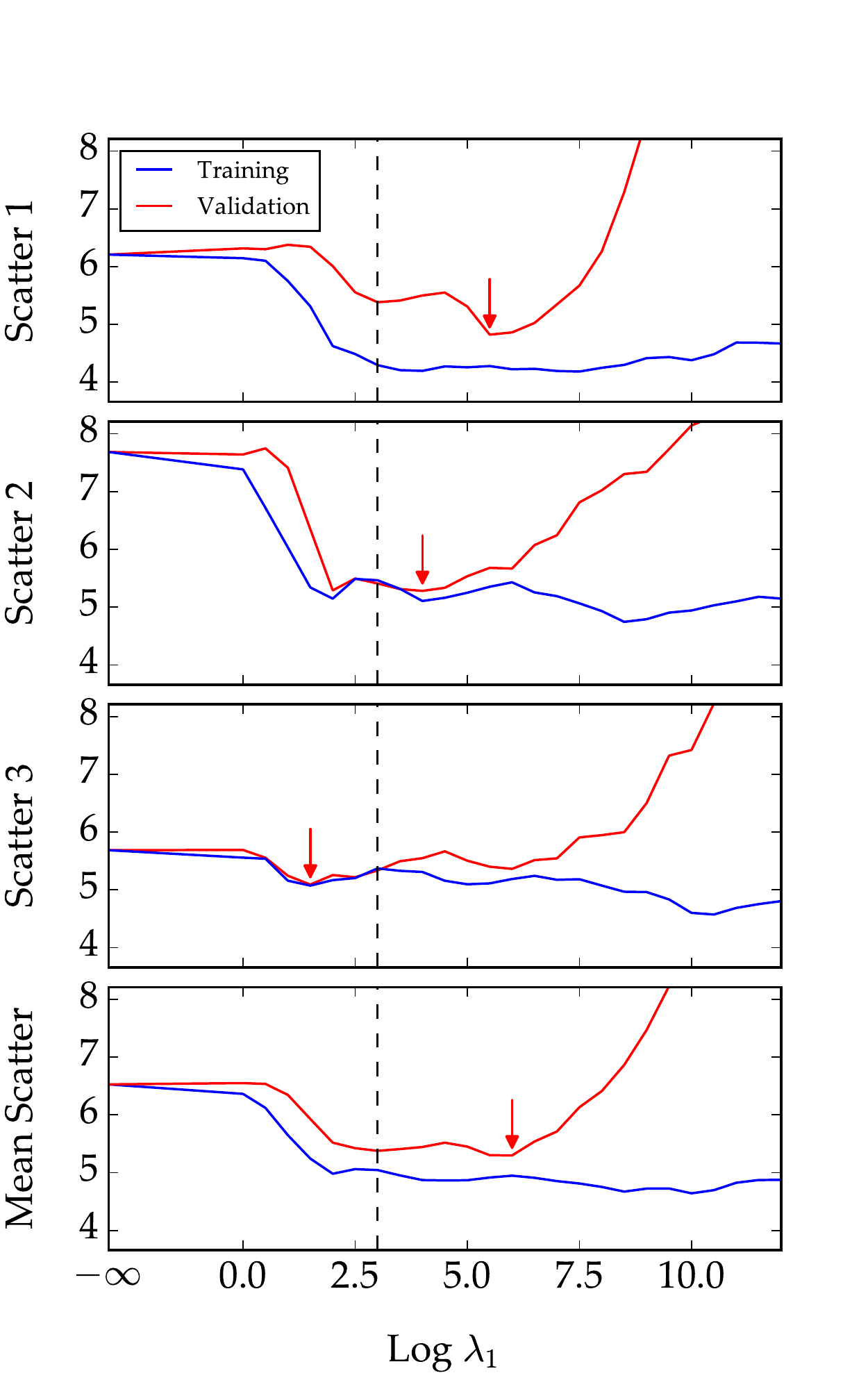}
       \caption{Cross-validation procedure for first order PLD on EPIC 206103150
       (WASP-47 e), a campaign 3 planet host. Shown is the scatter $\sigma_v$ in the validation
       set (red) and the scatter in the training set (blue) as a function of $\lambda_1$,
       the prior amplitude for the first order PLD weights, for each of three light curve
       sections; the mean scatter is shown at the bottom. Red arrows indicate the minima
       in the $\sigma_v$ curves for each section; note that because of variable noise
       properties across the campaign, they all occur at different values of $\lambda_1$.
       The dashed vertical line indicates the value of $\hat{\lambda}_n$ obtained by the
       procedure outlined in the text, which establishes a compromise between 
       slight underfitting in the first two segments and slight overfitting in the
       third.
       }
     \label{fig:crossval}
  \end{center}
\end{figure}

In addition to being more computationally tractable, there are two major benefits to
minimizing $\sigma_v$ in this fashion. First, since we perform cross-validation three times
(once for each PLD order), we are able to progressively refine the outlier masks (\S\ref{sec:impl_lightcurves}) and
the GP hyperparameters (\S\ref{sec:impl_gp}) for each target in between cross-validation steps. Second, it
allows for some leeway in how we determine the minimum validation scatter. In Paper I, 
we sought to minimize the \emph{median} scatter in groups of random 13-cadence segments 
of the light curve (the validation set). A potential issue with this method is that 
the noise properties of $K2$ light curves are far from constant over the course of
an observing campaign; optimizing the regression based on the median (or mean) validation 
scatter can still, in principle, lead to overfitting in some segments. While splitting the
light curves into segments with similar noise properties (\S\ref{sec:impl_breakpoints})
helps with this, we also modify the cross-validation process to prevent localized
overfitting. For each PLD order $n$ and for each value of $\lambda_n$, 
we split each light curve segment 
into three roughly equal sections. For each pair of sections, we train the model on
them and compute the model prediction in the third section (the validation set). We
then compute the scatter $\sigma_v$ as the median absolute deviation of the
de-trended validation set after removing the GP prediction.

We now have three $\sigma_v(\lambda_n)$ curves, one for each section. In general, the
minima of these curves will occur at different values of $\lambda_n$, so determining
the optimum value $\hat{\lambda}_n$ requires a compromise between overfitting and
underfitting in the different segments. For each segment, we compute the minimum scatter, 
find the set of all $\lambda_n$
for which $\sigma_v(\lambda_n)$ is within 5\% of the minimum, and keep the largest
$\lambda_n$. We then pick $\hat{\lambda}_n$ to be the \emph{smallest} of these
values, provided it is smaller than the largest value of $\lambda_n$ at the minima
of the three $\sigma_v$ curves. This process ensures that $\hat{\lambda}_n$ falls
between the minima of the $\sigma_v$ curves with the smallest and largest value of
$\lambda_n$, and that it leads to no more than 5\% overfitting in
one of the segments. We illustrate this procedure in Figure~\ref{fig:crossval}, where
we show $\sigma_v(\lambda_1)$ for each of the three light curve sections for
EPIC 206103150. The red arrows indicate the minimum of each of the curves, and the
dashed vertical line indicates the adopted $\hat{\lambda}_n$ based on a compromise
between slight underfitting in the first two segments and slight overfitting in the
third. This results in a more conservative cross-validation process than in Paper I.

\section{Results}
\label{sec:results}

We de-trended all campaigns 0--8 stars with \texttt{nPLD} to produce the 
\texttt{EVEREST 2.0} catalog. Below we report our results, starting with
injection/recovery tests and a comparison of \texttt{rPLD} and \texttt{nPLD},
followed by comparisons with other pipelines and the original \emph{Kepler}
light curves. We report most of our results in terms of the proxy 6 hr CDPP of
the de-trended light curves, which we calculate in the same way as we did
in Paper I: we smooth the light curves with a Savitsky-Golay filter, clip
outliers at 5$\sigma$, and compute the median standard deviation in 13-cadence
segments, normalized by $\sqrt{13}$.

\subsection{Injection Tests}
\label{sec:inj}

\begin{figure}[hbt]
  \begin{center}
      \includegraphics[width=0.47\textwidth]{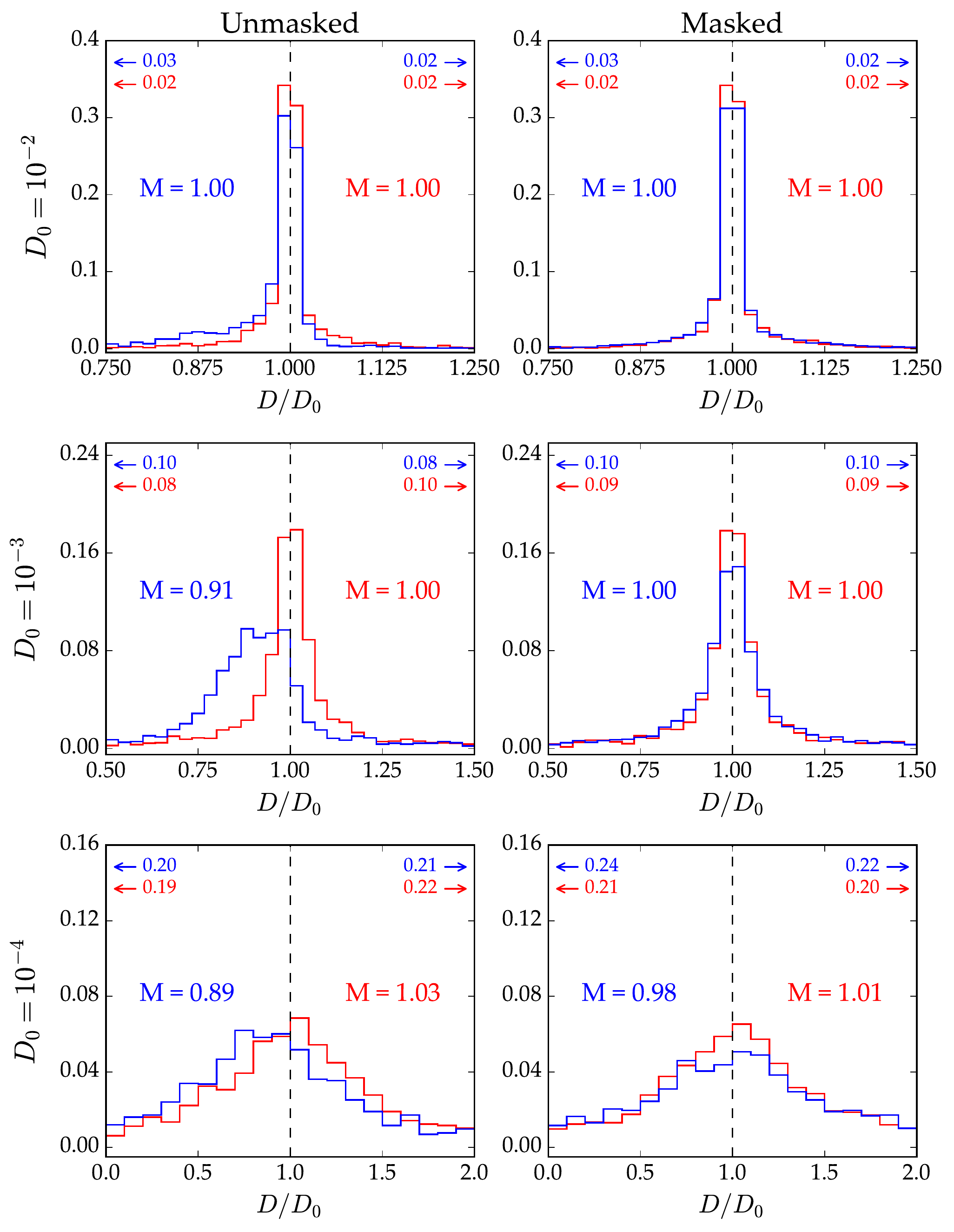}
       \caption{Transit injection/recovery statistics based on 2,700 randomly selected stars from campaign 6.
       Each panel shows histograms of the number of transits recovered with a certain depth 
       ratio $D/D_0$ (recovered depth divided by true depth). Blue histograms correspond to the actual
       injection and recovery process, in which transits are injected into the raw light curves at the pixel level
       and recovered after de-trending with \texttt{EVEREST}; red histograms correspond to control runs in which the transits
       were injected into the \emph{de-trended} data. The values 
       to the left and right of each histogram are the median $D/D_0$ for our pipeline and for the 
       control run, respectively. The smaller values at the top indicate the fraction of transits recovered 
       with depths lower and higher than the bounds of the plots. Finally, the two columns distinguish between 
       runs in which the transits were not explicitly masked prior to de-trending (left) and runs in which they were (right), 
       while the three rows correspond to different injected depths: $10^{−2}$, $10^{−3}$, and $10^{−4}$. \texttt{EVEREST} 
       preserves transit depths if the transits are properly masked; otherwise, a ${\sim}10\%$ bias toward smaller depths is 
       introduced for transits with low SNR.}
     \label{fig:injections}
  \end{center}
\end{figure}

As in Paper I, we perform simple transit injection/recovery tests to ensure
our model is not overfitting. For the same sample of 2,700 campaign 6 stars
as before, we injected synthetic transits of varying depths at the raw pixel level
and attempt to recover them after de-trending with \texttt{nPLD}. We follow
the exact same procedure as in \S4.1 of Paper I and plot the results in
Figure~\ref{fig:injections} (compare to Figure~6 in Paper I).

Each panel displays two histograms: a blue one, showing the number of
transits recovered with a certain depth after de-trending with \texttt{nPLD}, 
and a red one, corresponding to a control run in which the transits were injected
into the already de-trended light curve. Each row corresponds to a different
injection depth $D_0$ ($10^{-2}$, $10^{-3}$, and $10^{-4}$, from top to bottom), and the $x$ axis
in each histogram is the recovered depth $D$ scaled to this value ($D/D_0$). The 
left column corresponds to runs in which the transits were not explicitly masked
during de-trending; the right column shows runs in which they were.

As with the previous version of the pipeline, we find a ${\sim}10\%$ bias toward
smaller depths for low SNR transits when the transits are not explicitly masked. This
is because a small decrease in the transit depth can greatly improve the CDPP
of the light curve. Because the PLD regressors are noisy, the method is capable
of partially fitting out transits by exploiting linear combinations of white noise
in the regressors. This overfitting does not occur for high SNR transits because these
are masked during the outlier clipping step.

Conversely, when transits are explicitly masked, there is no bias in the recovered
depth; the median $D/D_0$ is consistent with unity for all three values of the injected 
depth. This is the same result we obtained with \texttt{EVEREST 1.0}, and we conclude
that our new cross-validation scheme is robust in preventing overfitting when transits
are masked. As before, we urge those using \texttt{EVEREST} light curves containing
transits or eclipses to re-compute the model with those features masked. This process
is quick and straightforward --- refer to the \texttt{EVEREST 2.0} documentation for
details.
\footnote{\url{https://github.com/rodluger/everest}}

\subsection{\texttt{rPLD}}
\begin{figure}[hbt]
  \begin{center}
      \includegraphics[width=0.47\textwidth]{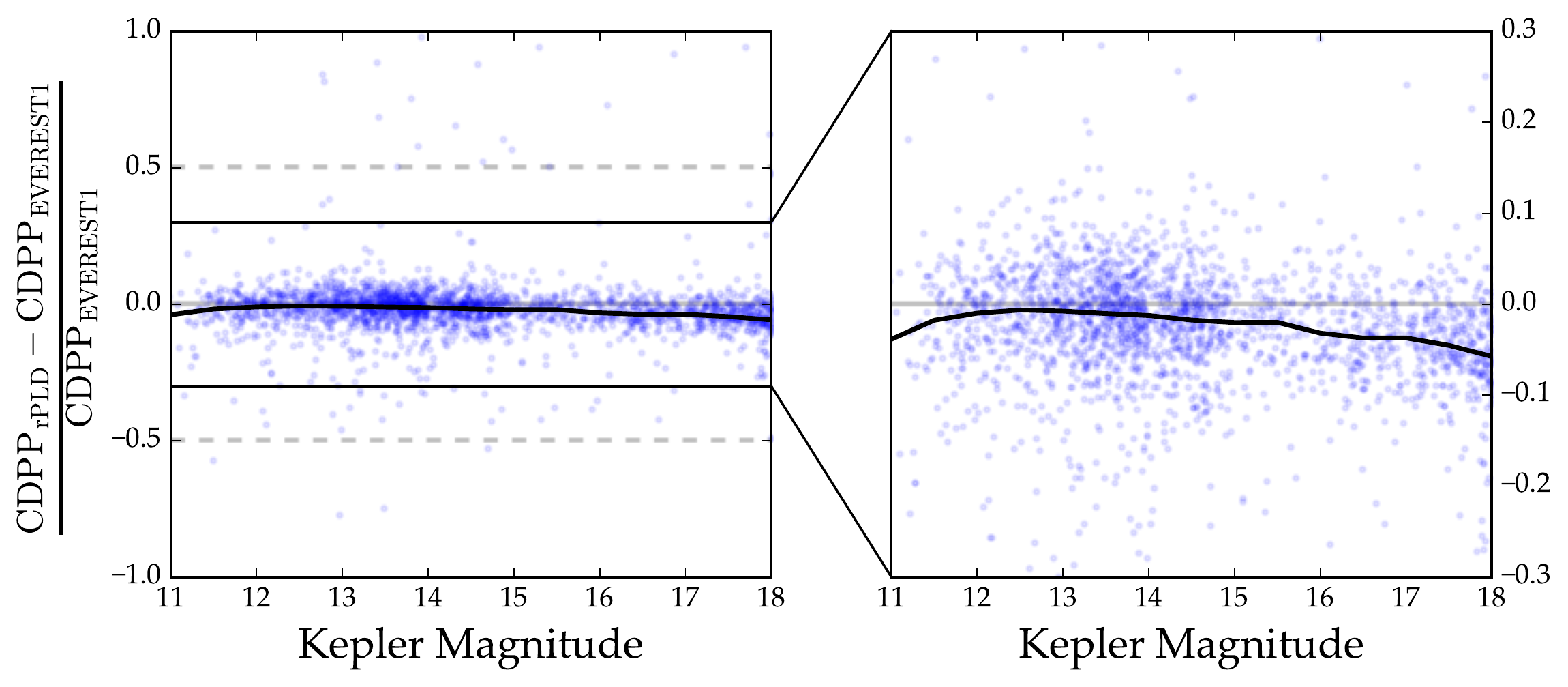}
       \caption{6 hr CDPP comparison between de-trending with regularized regression (\texttt{rPLD}, this paper) and
                de-trending with PCA (Paper I) for a sample of 2,700 randomly selected campaign 6
                stars, as in Figure~\ref{fig:pPLD}. Regularized regression leads to a small CDPP improvement of 
                ${\sim}1$ to 5\%.}
     \label{fig:rPLD}
  \end{center}
\end{figure}

In Figure~\ref{fig:rPLD} we plot a comparison of the CDPP values obtained with \texttt{rPLD}
and \texttt{EVEREST 1.0} for our sample set of 2,700 campaign 6 stars. As before, the $y$
axis corresponds to the normalized relative CDPP of each model, with negative values
corresponding to lower CDPP for \texttt{rPLD}. Each star is plotted as a blue dot and the median
relative CDPP is indicated as a black line. \texttt{rPLD} outperforms \texttt{EVEREST 1.0} at all
Kepler magnitudes by ${\sim}1-6\%$ on average. However, the scatter at any value of
$\Kp$ is quite large, and the two models are roughly comparable for bright stars. As we
argued in \S\ref{sec:impl_crossval}, the most important feature of \texttt{rPLD} is 
its increased robustness to local overfitting (see \S\ref{sec:remarks}).

\subsection{\texttt{nPLD}}
\begin{figure}[hbt]
  \begin{center}
      \includegraphics[width=0.47\textwidth]{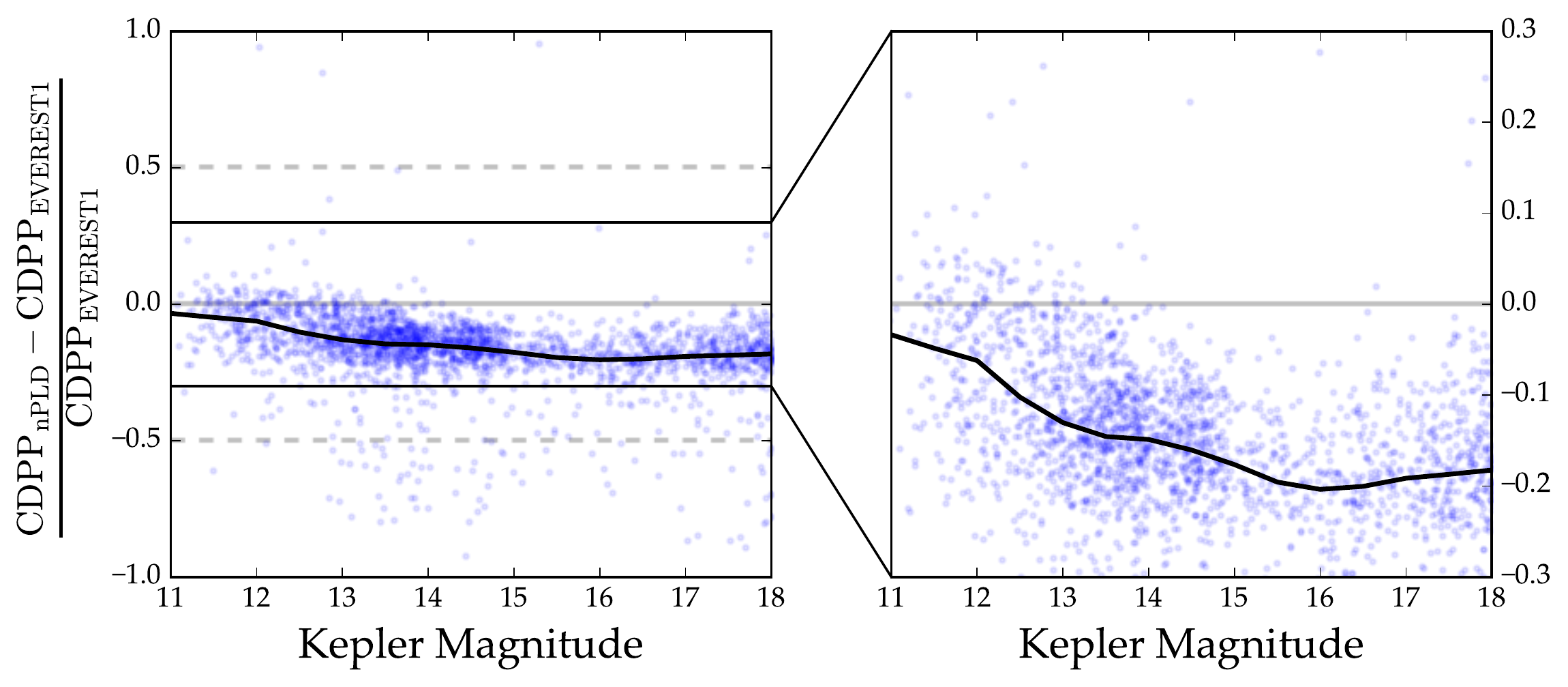}
       \caption{6 hr CDPP comparison between PLD de-trending with regularized regression + neighboring targets 
                (this paper) and standard PLD de-trending (Paper I) for the same sample of stars as in Figure~\ref{fig:rPLD}.
                Each target was de-trended with its own PLD vectors plus those of ten random bright stars
                on the same module. This method leads to a robust CDPP improvement of ${\sim}10$\% for bright ($\Kp \lesssim 13$) stars
                and ${\sim}20$\% for fainter stars.}
     \label{fig:nPLD}
  \end{center}
\end{figure}

The greatest improvement in the CDPP comes when neighboring stars' PLD vectors are included
in the design matrix. In Figure~\ref{fig:nPLD} we plot the CDPP comparison between \texttt{nPLD}
and \texttt{EVEREST 1.0}. \texttt{nPLD} outperforms regular PLD by ${\sim}10-20\%$
on average, with the largest improvement occurring for fainter stars. Faint stars have the
noisiest PLD vectors and benefit the most from the inclusion of higher SNR regressors.
As we showed in Paper I, regular PLD already comes close to recovering \emph{Kepler} photometric 
precision for bright ($\Kp \lesssim 13$) stars, so the improvement for these stars is naturally 
smaller.

\subsection{Comparison to Other Pipelines}
\label{sec:comparison}

\begin{figure}[hbt]
  \begin{center}
      \includegraphics[width=0.47\textwidth]{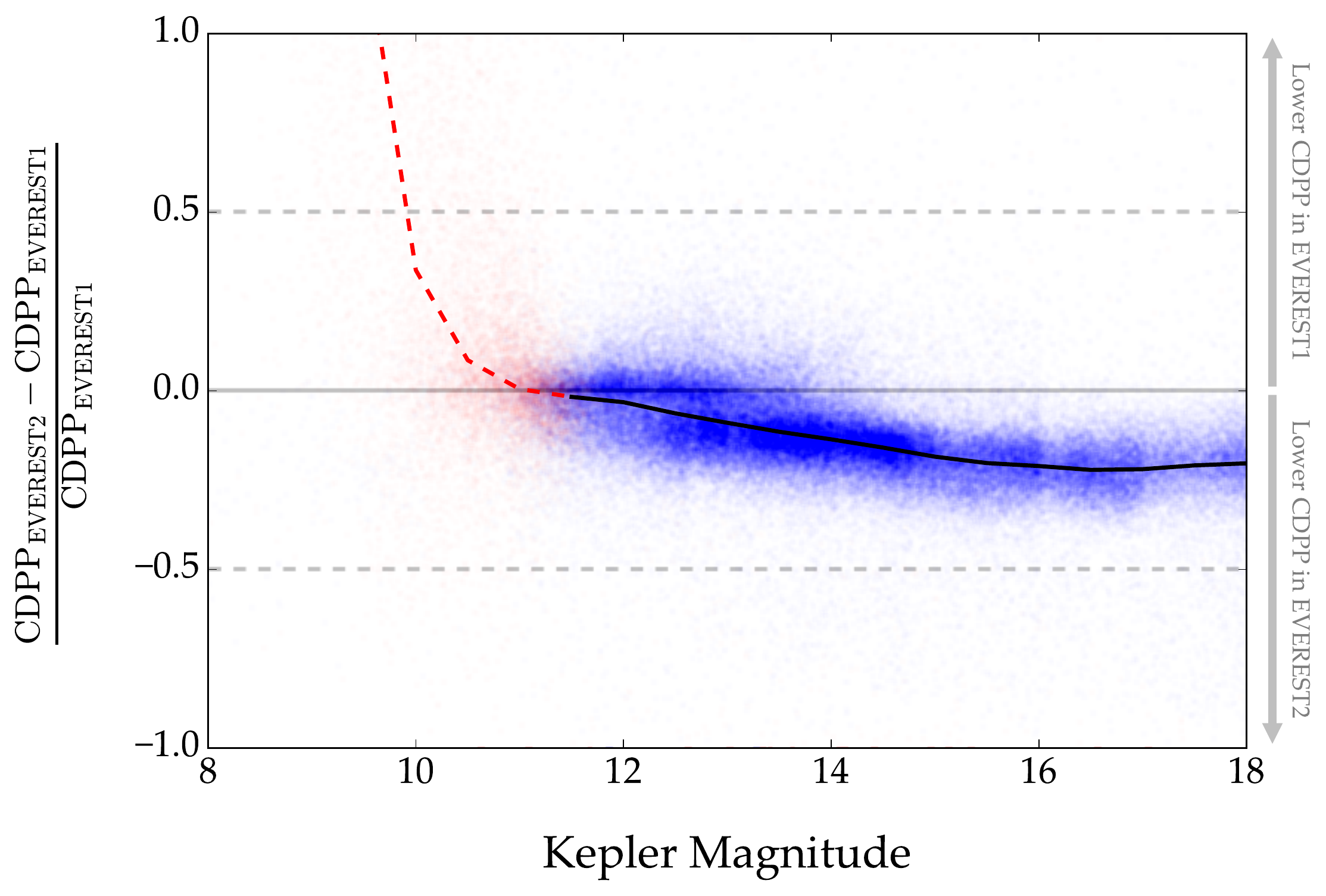}
       \caption{CDPP comparison between \texttt{EVEREST 2.0} (using \texttt{nPLD}) and \texttt{EVEREST 1.0} for all stars in campaigns 0--8.
       As before, individual stars are plotted as blue points and the median CDPP is indicated by a black line; note
       the ${\sim}10-20$\% improvement over the previous version of the pipeline. Saturated
       stars are plotted as red points, with their median CDPP indicated by a dashed red line. The apparently
       better performance of \texttt{EVEREST 1.0} for these stars is spurious, since traditional PLD typically
       leads to strong overfitting of saturated stars (see Figure~\ref{fig:saturated_star}).}
     \label{fig:cdpp_everest1_all}
  \end{center}
\end{figure}

In this section we compare the \texttt{EVEREST 2.0} catalog to those produced by other pipelines, beginning with the
previous version of \texttt{EVEREST}.

Figure~\ref{fig:cdpp_everest1_all} shows the CDPP comparison between \texttt{EVEREST 2.0} and
\texttt{EVEREST 1.0} for all campaigns 0--8 stars. As in the example shown in Figure~\ref{fig:nPLD},
\texttt{EVEREST 2.0} outperforms \texttt{EVEREST 1.0} by ${\sim}20\%$ for the faintest stars and by
${\sim}10\%$ for $\Kp \gtrsim 12$. For $11 \lesssim \Kp \lesssim 12$ the two pipelines yield
comparable results, though regularized regression gives \texttt{EVEREST 2.0} a slight edge. Below
$\Kp \approx 11$, $K2$ stars become saturated; these are plotted as red dots, and the median
relative CDPP for saturated stars is indicated by the dashed red line. For these stars,
\texttt{EVEREST 1.0} yields much lower CDPP --- for $\Kp \lesssim 10$, the CDPP is over a factor
of two smaller than that of \texttt{EVEREST 2.0}. As we discussed in Paper I, the increased
performance of \texttt{EVEREST 1.0} for saturated stars is spurious, since the astrophysical
information content of the pixels is highly variable across the aperture, leading regular
PLD to overfit. As we showed in \S\ref{sec:impl_saturated}, \texttt{EVEREST 2.0} does not 
overfit saturated stars. In \S\ref{sec:kepler} below, we show that we approximately recover
the \emph{Kepler} photometric precision for these stars.

\begin{figure}[hbt]
  \begin{center}
      \includegraphics[width=0.47\textwidth]{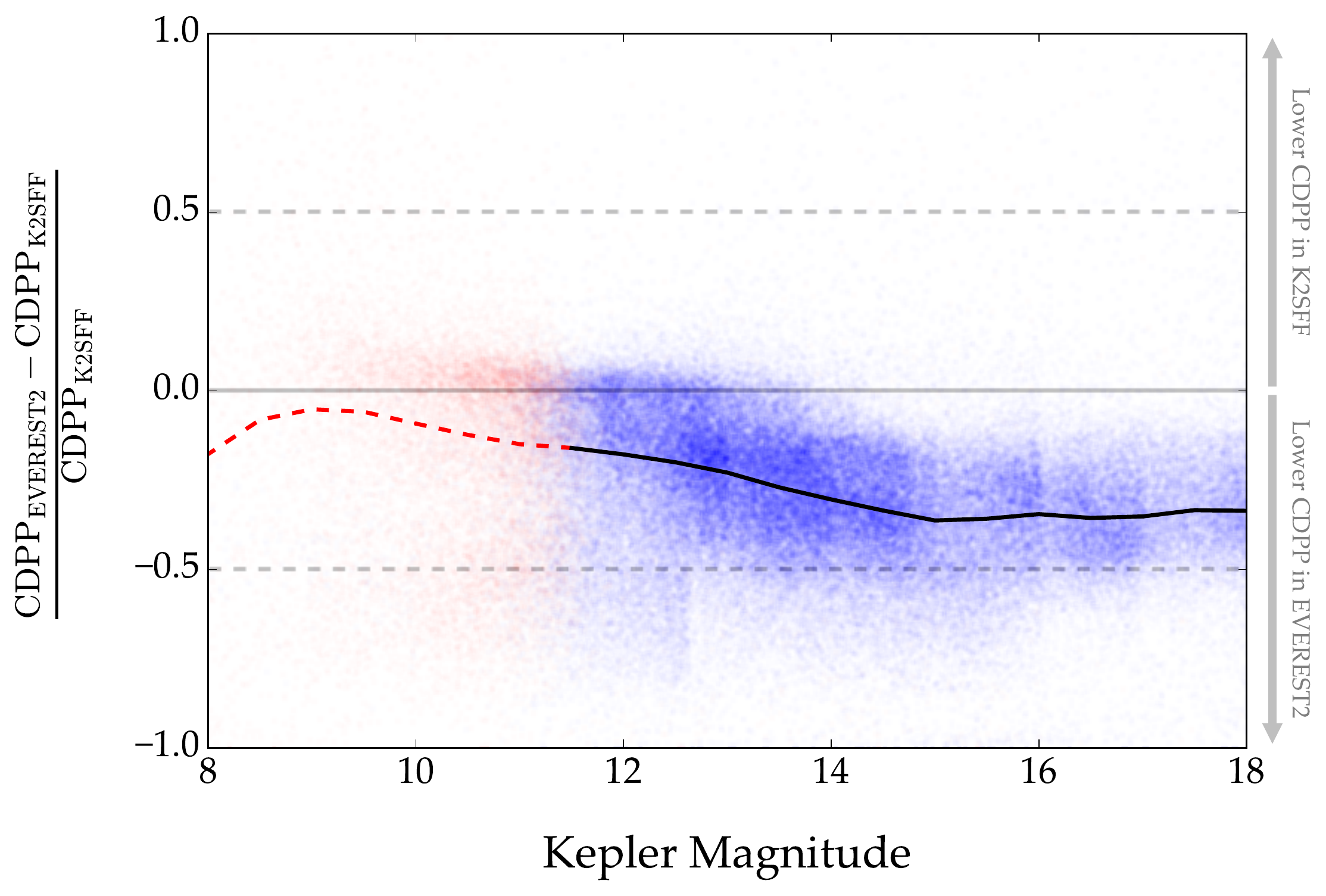}
       \caption{Similar to Figure~\ref{fig:cdpp_everest1_all}, but showing a comparison between \texttt{EVEREST 2.0} and \texttt{K2SFF}.
       \texttt{EVEREST 2.0} outperforms \texttt{K2SFF} at all magnitudes, including $\Kp \lesssim 11$, for which stars are saturated.}
     \label{fig:cdpp_k2sff_all}
  \end{center}
\end{figure}

In Figure~\ref{fig:cdpp_k2sff_all} we show the CDPP comparison between \texttt{EVEREST 2.0}
and \texttt{K2SFF} \citep{Vanderburg14,VanderburgJohnson14}. Our pipeline yields
lower average CDPP at all magnitudes, by 40\% for faint stars and 10--20\% for bright
stars. For saturated stars, \texttt{EVEREST 2.0} outperforms \texttt{K2SFF} by 5--10\%.

\begin{figure}[hbt]
  \begin{center}
      \includegraphics[width=0.47\textwidth]{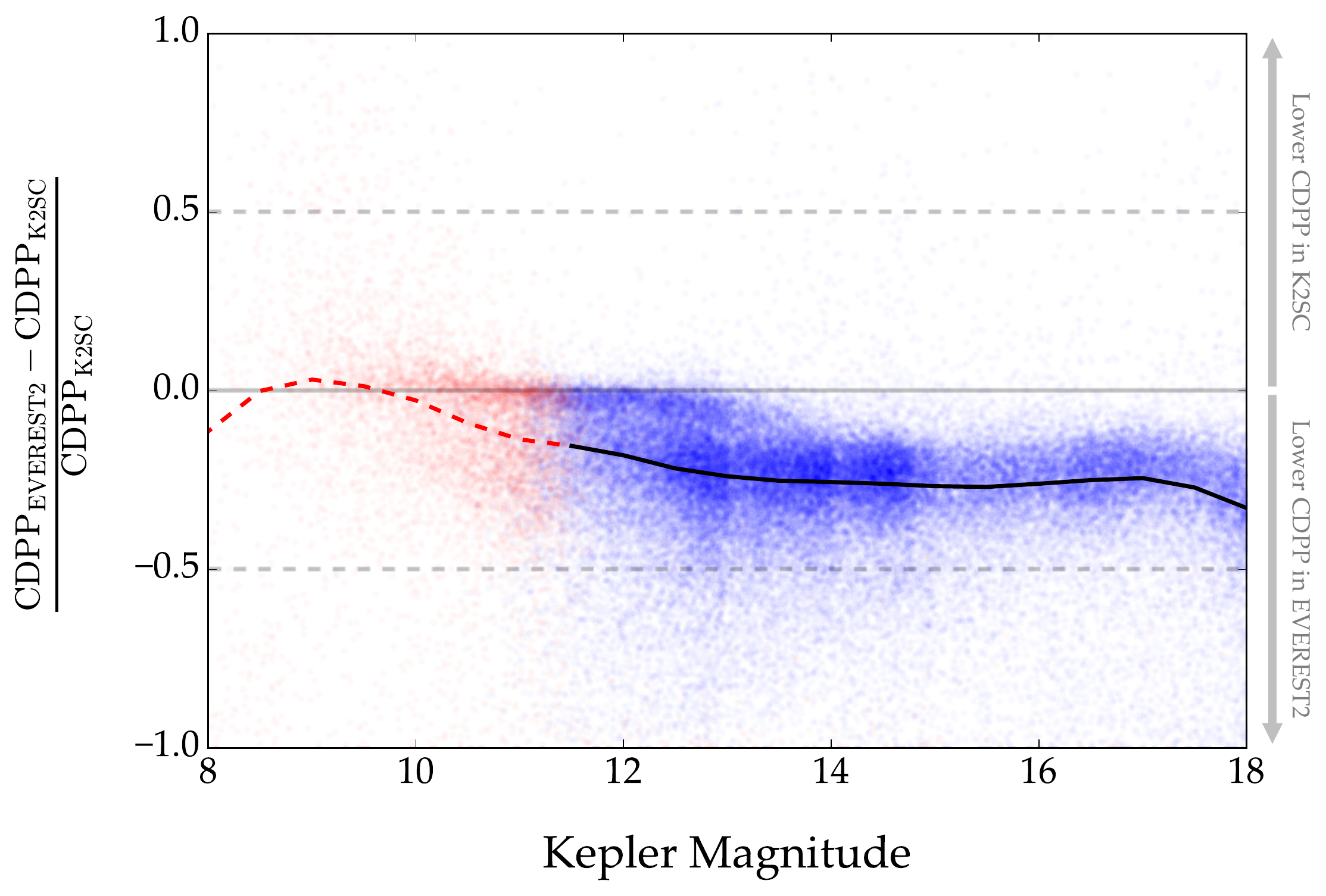}
       \caption{Similar to the previous figure, but showing a comparison between \texttt{EVEREST 2.0} and \texttt{K2SC}.
       \texttt{EVEREST 2.0} light curves have lower average CDPP at all magnitudes except around $\Kp \approx 9$, for which
       the precision is comparable.}
     \label{fig:cdpp_k2sc_all}
  \end{center}
\end{figure}

In Figure~\ref{fig:cdpp_k2sc_all} we show the comparison to the \texttt{K2SC} PDC
light curves \citep{Aigrain15,Aigrain16}. \texttt{EVEREST 2.0} yields lower average CDPP at all
magnitudes $K_p \gtrsim 9$, with a 20--25\% improvement for
all unsaturated stars. For $K_p \approx 9$, \texttt{K2SC} slightly overperforms
\texttt{EVEREST 2.0}, but the scatter in the plot is quite large.

\begin{figure}[hbt]
  \begin{center}
      \includegraphics[width=0.47\textwidth]{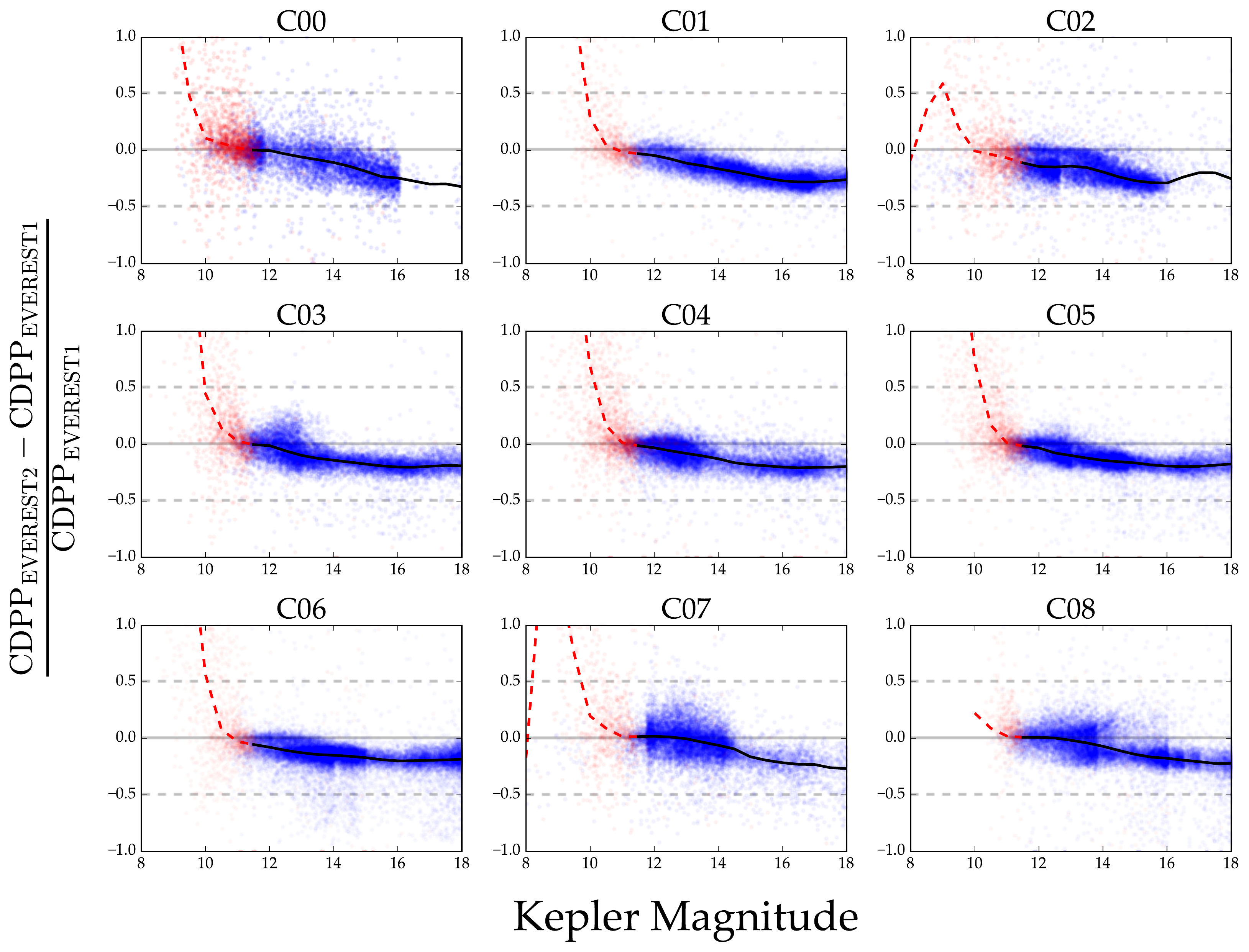}
       \caption{Similar to Figure~\ref{fig:cdpp_everest1_all}, but showing a CDPP comparison between 
       \texttt{EVEREST 2.0} and \texttt{EVEREST 1.0} for each of the first 9 $K2$ campaigns.}
     \label{fig:cdpp_everest1}
  \end{center}
\end{figure}

\begin{figure}[hbt]
  \begin{center}
      \includegraphics[width=0.47\textwidth]{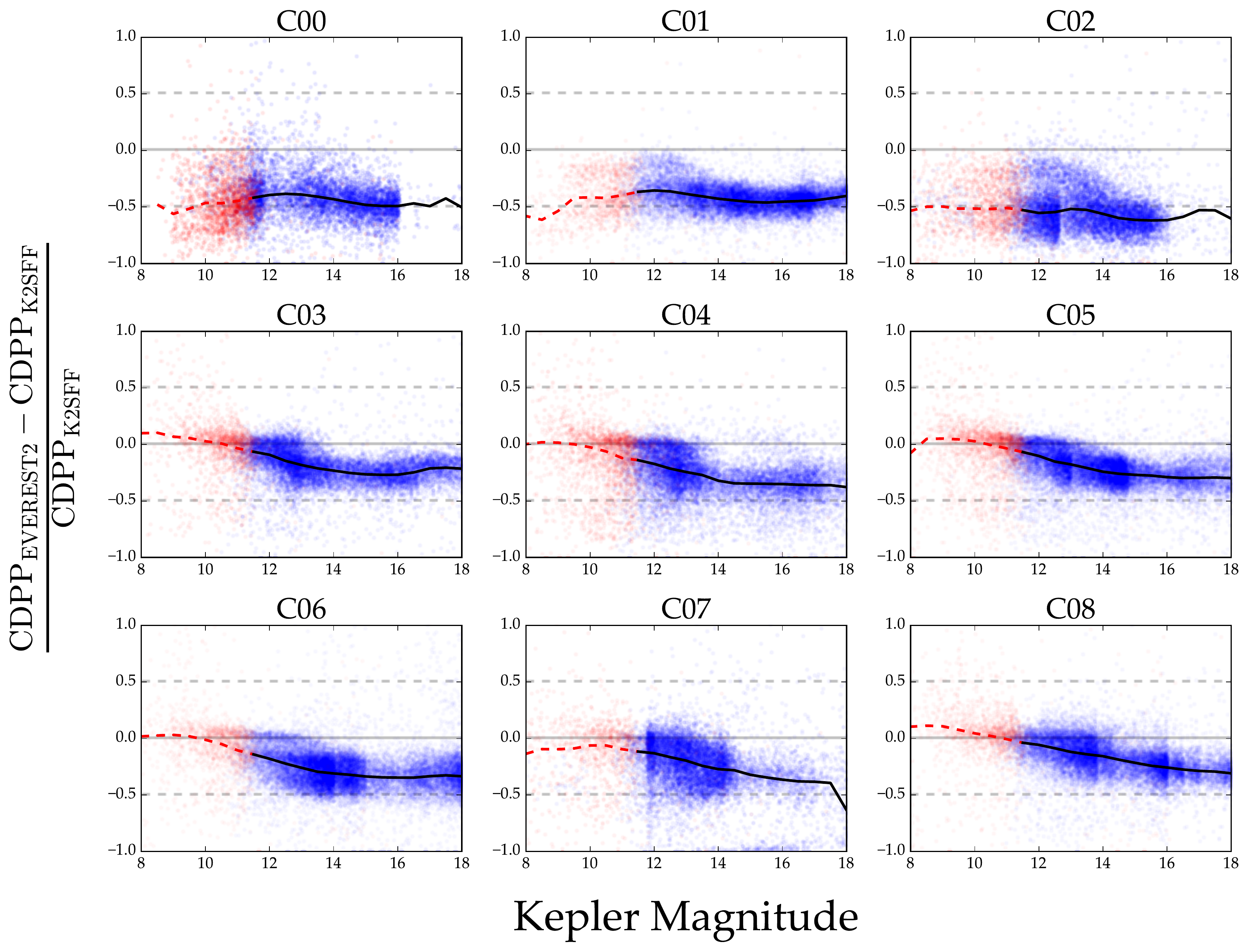}
       \caption{Similar to Figure~\ref{fig:cdpp_k2sff_all}, but showing a CDPP comparison between 
       \texttt{EVEREST 2.0} and \texttt{K2SFF} for each of the first 9 $K2$ campaigns.}
     \label{fig:cdpp_k2sff}
  \end{center}
\end{figure}

\begin{figure}[hbt]
  \begin{center}
      \includegraphics[width=0.47\textwidth]{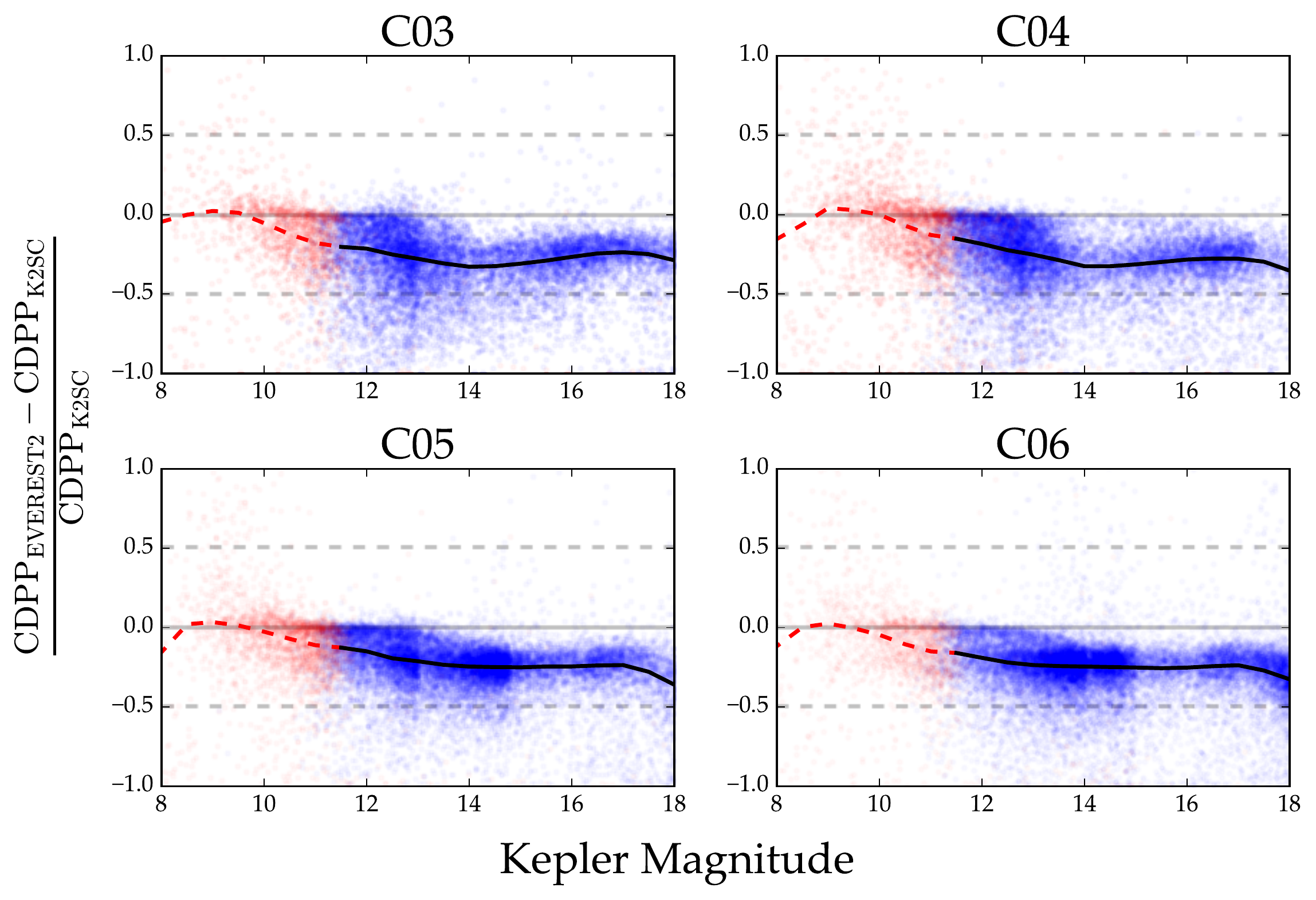}
       \caption{Similar to Figure~\ref{fig:cdpp_k2sc_all}, but showing a CDPP comparison between 
       \texttt{EVEREST 2.0} and \texttt{K2SC} for each of the first 9 $K2$ campaigns.}
     \label{fig:cdpp_k2sc}
  \end{center}
\end{figure}

We also computed the relative CDPP for each of the campaigns individually; these are
plotted in Figures~\ref{fig:cdpp_everest1}--\ref{fig:cdpp_k2sc}. The improvement
over \texttt{EVEREST 1.0} is approximately the same for all campaigns despite significant
differences in the noise properties and stellar populations across the nine campaigns,
showcasing the robustness of the \texttt{nPLD} method.
The same is true when compared to \texttt{K2SFF} and \texttt{K2SC}, with the exception of
campaigns 0--2, for which \texttt{EVEREST 2.0} outperforms \texttt{K2SFF} by nearly 50\%
(i.e., a factor of 2) at all magnitudes.

\subsection{Outliers}

\begin{figure}[hbt]
  \begin{center}
      \includegraphics[width=0.47\textwidth]{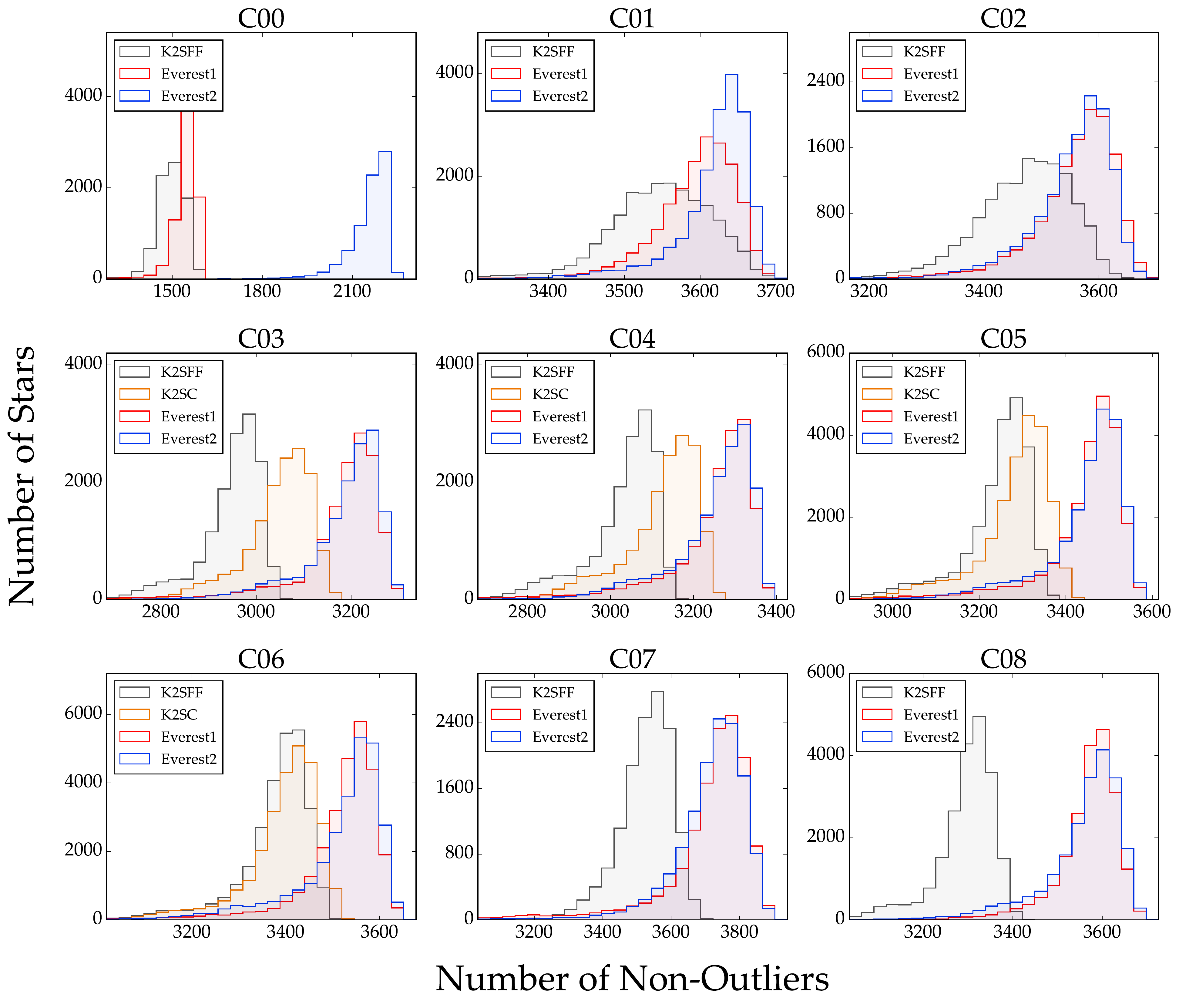}
       \caption{Histograms showing the number of non-outlier data points per campaign for each
        of four pipelines: \texttt{K2SFF} (gray), \texttt{K2SC} (orange; campaigns 3--6 only), 
        \texttt{EVEREST 1.0} (red), and \texttt{EVEREST 2.0} (blue). To compute these, we remove all cadences 
        with flagged \texttt{QUALITY} bits (excepting thruster fires) from all light curves, then
        smooth each light curve with a second order, 2-day Savitsky-Golay filter and perform iterative 
        sigma clipping at 5$\sigma$ to remove the outliers. The number of remaining cadences in
        each light curve is then used to plot the histograms. Both versions of \texttt{EVEREST}
        have more usable data points per campaign than the other pipelines. On average, 
        \texttt{EVEREST} light curves have ${\sim}200-300$ more non-outlier data points than \texttt{K2SFF} 
        and ${\sim}100$ more than \texttt{K2SC}.}
     \label{fig:outliers}
  \end{center}
\end{figure}

In addition to yielding lower average CDPP than any other publicly available pipeline,
\texttt{EVEREST} yields the largest number of \emph{usable} data points for any of the
$K2$ campaigns to date. These data points generally appear as outliers even in the
raw data, since they are highly sensitive to inter- and intra-pixel sensitivity variations.
Pipelines that regress on functions of the spacecraft motion alone therefore have trouble
de-trending them, resulting in ${\sim}5-10\%$ of the data points being discarded as outliers.
In contrast, since PLD uses regressors containing both inter- and intra-pixel sensitivity
information, it naturally de-trends data collected during thruster firing events (see Paper I).

To show this, we calculated the number of non-outlier data points per campaign
for each of the four pipelines (\texttt{K2SFF}, \texttt{K2SC}, \texttt{EVEREST 1.0}, and
\texttt{EVEREST 2.0}). After removing all data points with flagged \texttt{QUALITY}
bits 1-9, 11-14, and 16-17, we smoothed each light curve with a Savitsky-Golay filter and
performed iterative sigma clipping to remove all 5$\sigma$ outliers. In Figure~\ref{fig:outliers}
we plot histograms of the number of remaining data points for all stars in each of the first
9 campaigns. As expected, both \texttt{EVEREST 1.0} and \texttt{EVEREST 2.0} have, on 
average, 100--300 more usable data points than the other two pipelines. This roughly
corresponds to the number of thruster firings per campaign, as these happen every 6--12
hours on average.

\subsection{Short Cadence}
\label{sec:shortcad}

\begin{figure}[hbt]
  \begin{center}
      \includegraphics[width=0.47\textwidth]{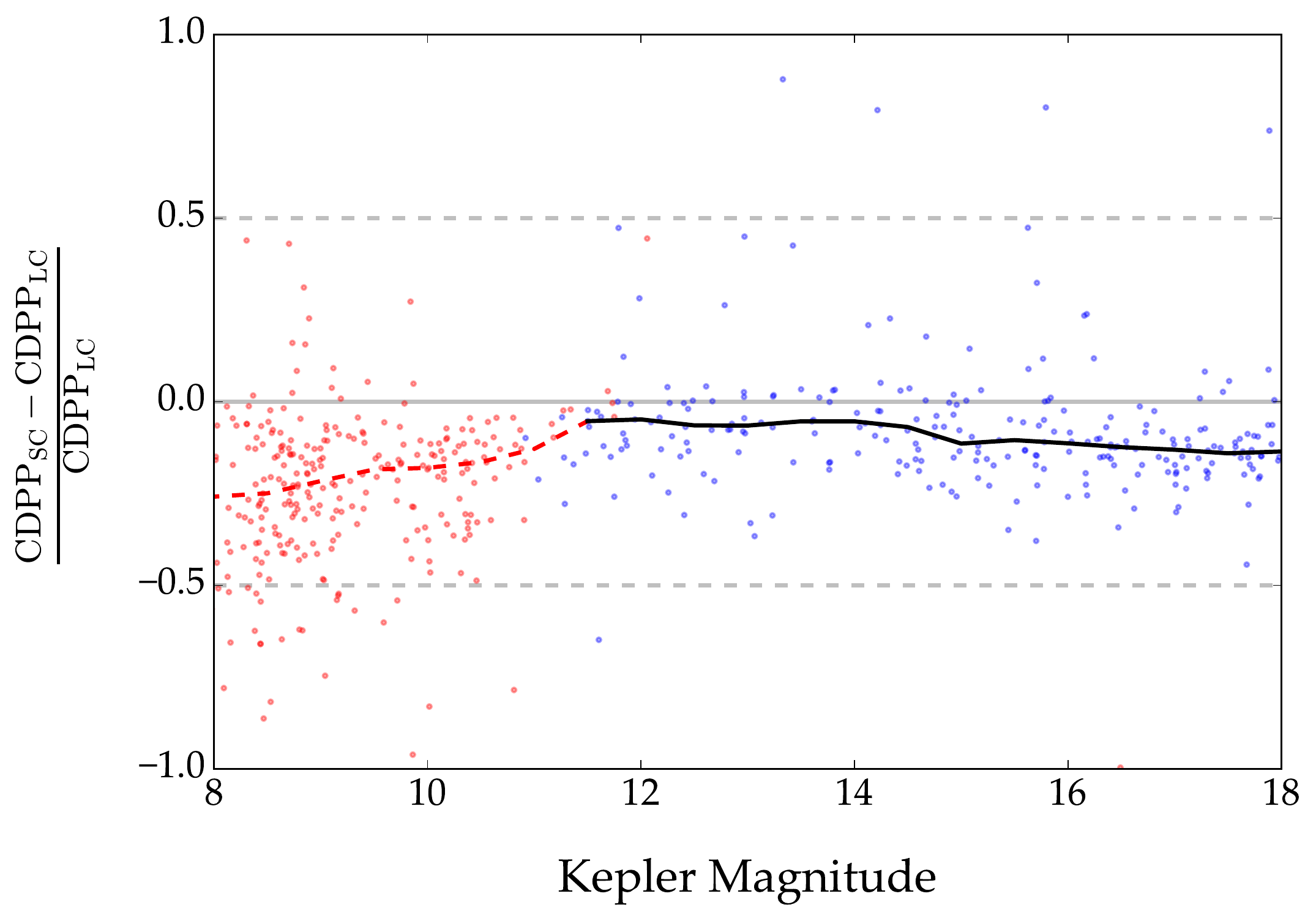}
       \caption{CDPP comparison between \texttt{EVEREST 2.0} light curves
       observed in short cadence and long cadence modes. Short cadence
       light curves have 5-10\% lower CDPP on average; for saturated stars,
       short cadence light curves have up to 25\% lower CDPP.}
     \label{fig:short_cad_stats}
  \end{center}
\end{figure}

\begin{figure}[hbt]
  \begin{center}
      \includegraphics[width=0.47\textwidth]{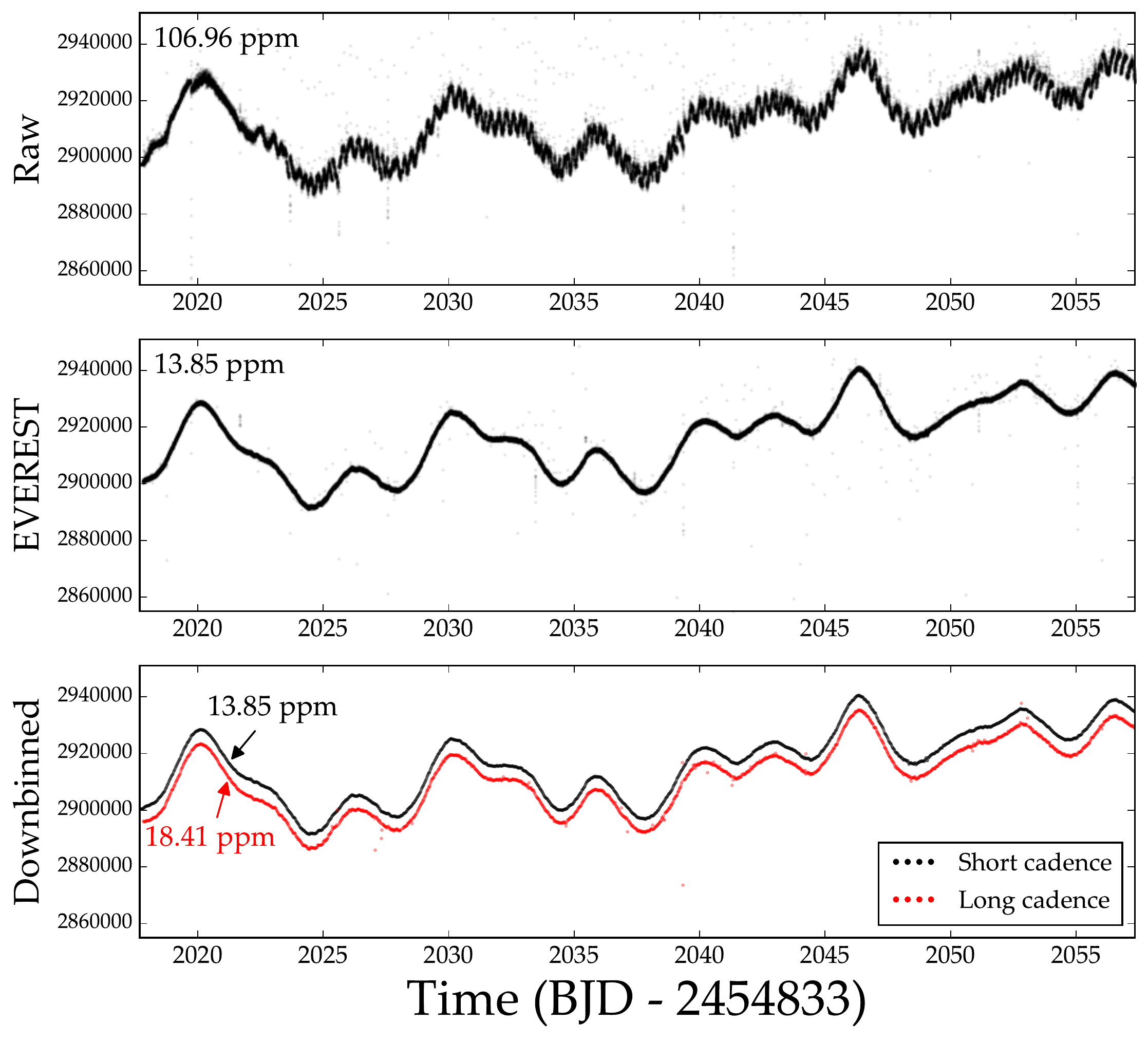}
       \caption{EPIC 201601162, a campaign 1 star observed in both long cadence and
       short cadence modes. A portion of the raw light curve is 
       displayed at the top, and the de-trended light curve is shown in the center. In the bottom panel,
       we plot the down-binned de-trended short cadence light curve (black) and the
       de-trended long cadence light curve (red). The $y$ axis in each panel is the flux
       in e$^{-}$/s.
       Short cadence \texttt{EVEREST 2.0}
       light curves have lower CDPP than their long cadence counterparts.}
     \label{fig:short_cad}
  \end{center}
\end{figure}

In Figure~\ref{fig:short_cad_stats} we plot the relative CDPP distribution of 
671 light curves that were observed in both short and long cadence. In order to
compute the 6 hr CDPP of short cadence light curves, we first mask outliers and then
down-bin to long cadence by taking the mean of every 30 cadences; we then compute
the CDPP as usual. We achieve higher average precision in the short cadence light
curves by 5-10\% for unsaturated stars and by up to 25\% for saturated stars. The
higher information content in the short cadence light curves --- particularly
at sub-30 minute timescales --- allows \texttt{EVEREST} to better de-trend those stars.
We show an example in Figure~\ref{fig:short_cad}, where we plot the light curves
for EPIC 201601162 (raw short cadence at the top, de-trended short cadence in the
center). In the bottom panel, we show both the down-binned short cadence de-trended light
curve (black) and the long cadence de-trended light curve (red). The CDPP in the short
cadence light curve is ${\sim}30\%$ smaller than that of the long cadence light curve.

As we discussed in \S\ref{sec:impl_shortcad}, one pitfall of our method for de-trending
short cadence targets is the large number of breakpoints (${\sim}30$) we introduce
in the light curve when computing the model. Overcomputing the model into adjacent
segments and aligning the models at the breakpoints works very well for high
SNR light curves but can often fail for light curves that are dominated by photon
noise. This is the case for very faint stars ($\Kp \gtrsim 17$) observed in
short cadence mode, such as EPIC 201831393, which displays visible discontinuities
at many of the breakpoints. Low SNR short cadence light curves may also have segments 
with varying white noise amplitudes due to differences in the value of the $\mathbf{\Lambda}$
prior on the PLD weights. PLD is known to perform poorly when the white noise dominates
\citep{Deming15}; in these cases, it is often desirable to down-bin the light curve and
compute the model on a higher SNR signal, then predict onto the original short cadence
data. Given the relatively small number of light curves for which this is an issue,
we do not do this here.

\subsection{Co-trending Basis Vectors}
\label{sec:cbvs}
One downside of the algorithm employed by \texttt{EVEREST} is that GP regression has trouble
distinguishing between low frequency stellar variability and low frequency instrumental
systematics. Many of the \texttt{EVEREST} light curves display a steady rise over the
course of the campaign, with hook-like features at the beginning, end, or both. While this
does not affect transits or other high-frequency astrophysical signals, it could potentially
lead to biases in stellar rotation studies that rely on low frequency modulation in the
light curves.

We therefore run a post-processing step on all de-trended light curves to remove these
residual instrumental signals. After some experimentation, we decided to use the
\texttt{SysRem} algorithm \citep{Tamuz05}, which identifies and removes signals shared by many
light curves. \texttt{SysRem} is similar to PCA but allows for weighting of the input
signals and is therefore better suited to dealing with light curves of varying noise
properties. Our approach is similar to the presearch data conditioning (PDC) algorithm
of the \emph{Kepler} pipeline, which uses co-trending basis vectors from many stars
on the detector to remove common instrumental signals from light curves \citep{Stumpe12,Smith12}.

\begin{figure}[hbt]
  \begin{center}
      \includegraphics[width=0.4\textwidth]{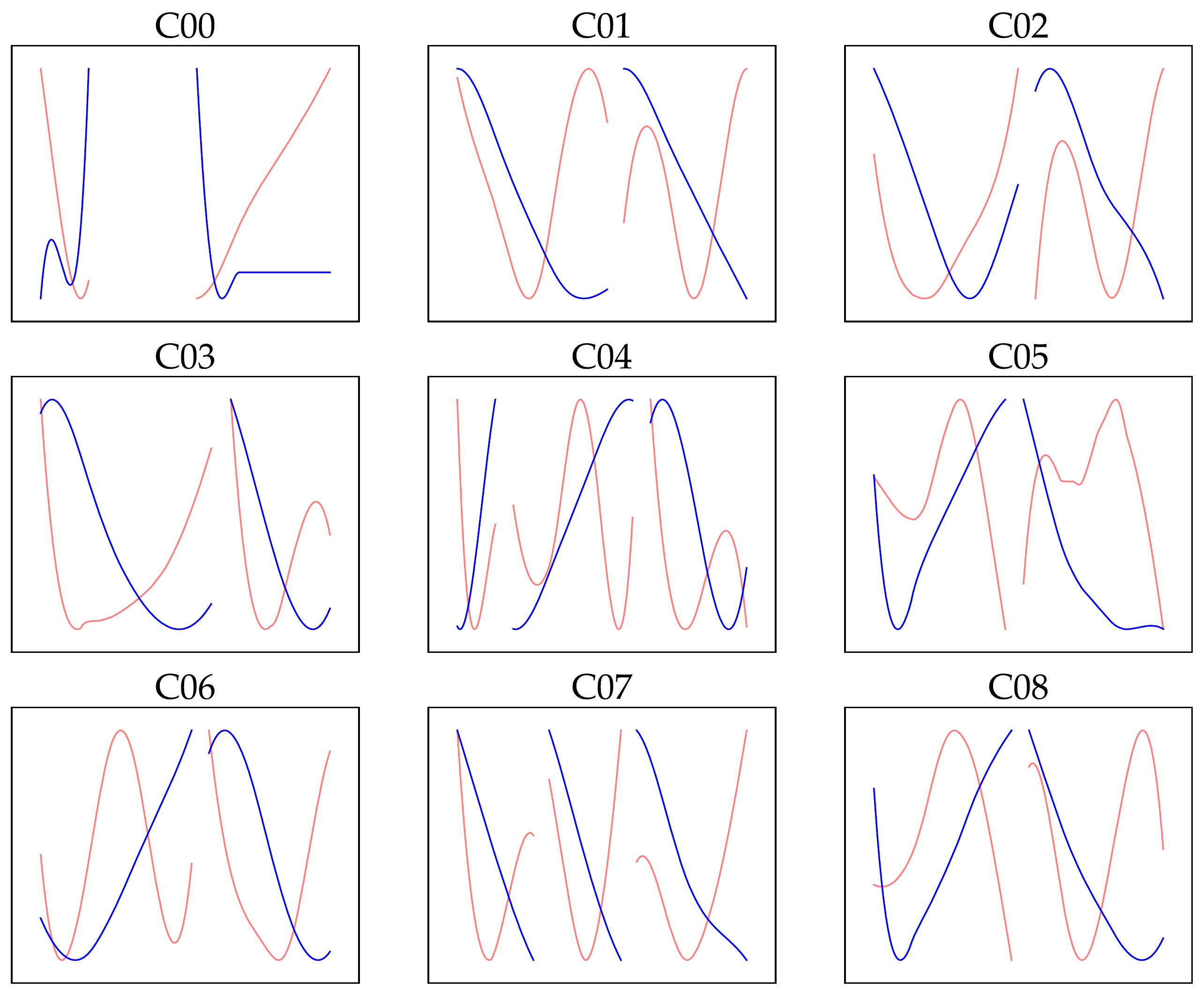}
      \caption{Co-trending Basis Vectors (CBVs) for each of the first 9 campaigns. 
         We apply \texttt{SysRem} to all de-trended light curves 
         in each campaign to obtain the first (blue) and second (red) CBVs; we do this
         independently for each of the segments in each campaign.
         The first set of CBVs contain primarily linear trends with hook-like
         features at the beginning or end of the segments; the second set of CBVs are
         dominated by quadratic or cubic trends. We correct all 
         light curves by simple
         linear regression with the first two CBVs.}
     \label{fig:cbv}
  \end{center}
\end{figure}

We apply \texttt{SysRem} to all of the light curves in each campaign, weighting each one
by the quadrature sum of the flux measurement errors and the white noise component of its
GP. Given known correlations between instrumental signals and spatial position on the
detector \citep[e.g.,][]{Petigura12,Wang16}, we attempted to apply \texttt{SysRem} individually
on each CCD module, but found that the recovered signals were often dominated by
astrophysical variability originating from the brightest star(s) in each module. 
We found that computing the \texttt{SysRem} signals for the entire detector alleviated 
this issue without compromising the de-trending power of the method.

We separately apply \texttt{SysRem} to each light curve segment, obtaining one
co-trending basis vector (CBV) for each segment of each campaign. We then subtract a linear
fit of this CBV from each light curve and repeat the procedure to obtain additional CBVs
for each segment of each campaign. In order to prevent the CBVs from fitting out or
introducing high frequency signals in the light curves, we aggressively smooth them
with a third order, 1000-cadence Savitsky-Golay filter. The results are shown in
Figure~\ref{fig:cbv}. The first CBV for each segment of each campaign is plotted in blue,
and the second in red. As expected, the first CBV in each segment is dominated by a linear trend with S-like
hooks on either end. The second CBVs are predominantly quadratic or cubic. The remaining
CBVs (not shown) are dominated by higher order trends.

For the purposes of generating the \texttt{EVEREST 2.0} catalog,
we perform ordinary least squares regression to fit all de-trended light curves
using only the first CBV (blue curves in the figure). We find that fitting with additional CBVs often helps
to remove additional systematics --- in particular the hook-like features mentioned above
--- but may lead to overfitting of true astrophysical variability in some light curves.
We therefore include all 5 CBVs of each segment of each campaign in the \texttt{EVEREST 2.0} \texttt{FITS} files
so that targets may be further corrected by the user on a case-by-case basis (see \S\ref{sec:using}).

%We also attempted to regress using the GP of each target, but there is 
%typically enough power in the GP model to strongly disfavor a CBV fit. In practice,
%our scheme works surprisingly well. 

\subsection{Sample Light Curves}
\label{sec:sample}

\begin{figure}[hbt]
  \begin{center}
      \includegraphics[width=0.47\textwidth]{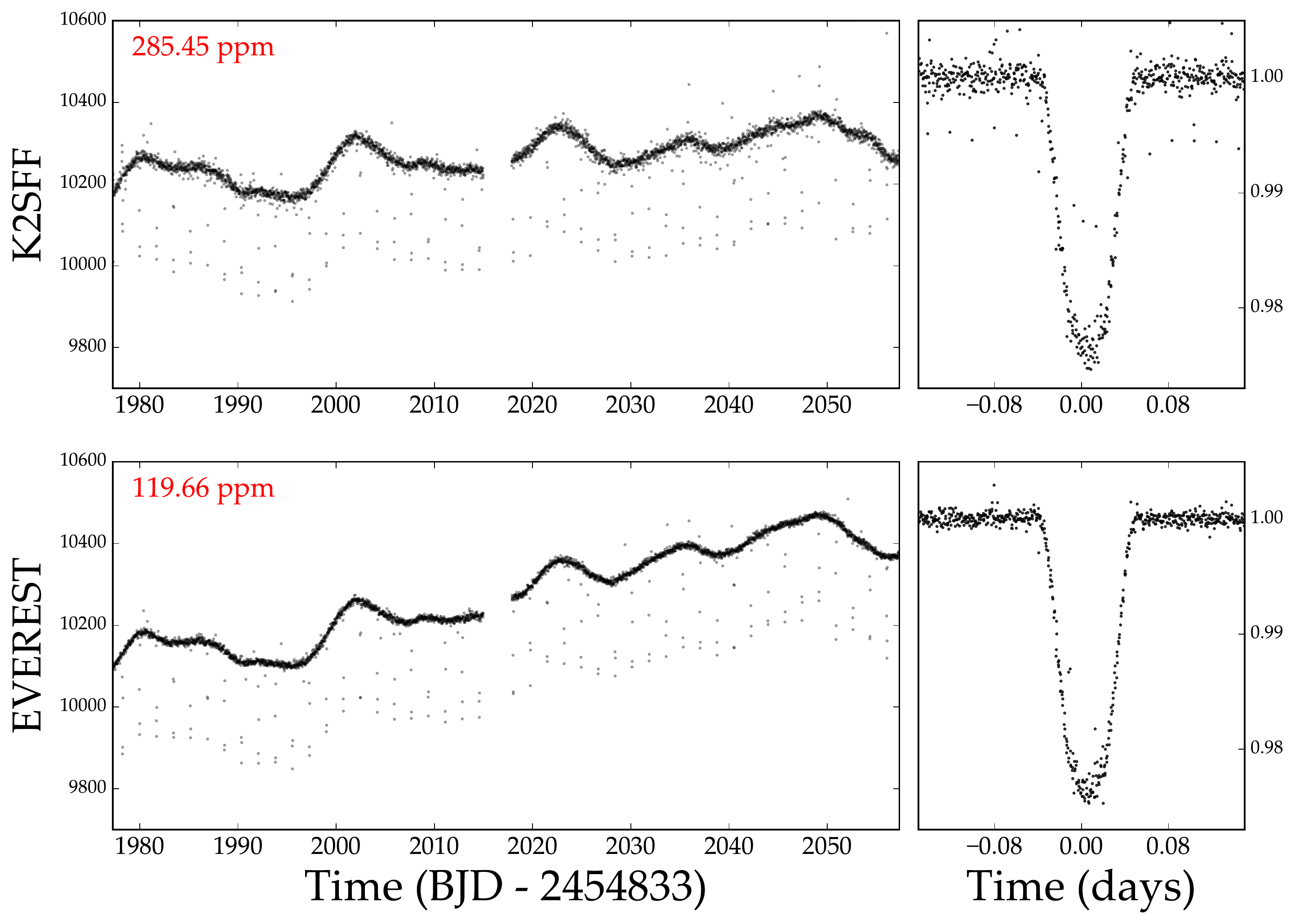}
       \caption{EPIC 201345483 (K2-45), a $\Kp = 15$ campaign 1 planet host de-trended
       with \texttt{K2SFF} (top) and \texttt{EVEREST 2.0} (bottom). The CDPP of each light curve
       is indicated in the top left. The folded transit of K2-45b is shown at right. The \texttt{EVEREST 2.0}
       light curve has 2.4$\times$ higher photometric precision.}
     \label{fig:201345483}
  \end{center}
\end{figure}

\begin{figure}[hbt]
  \begin{center}
      \includegraphics[width=0.47\textwidth]{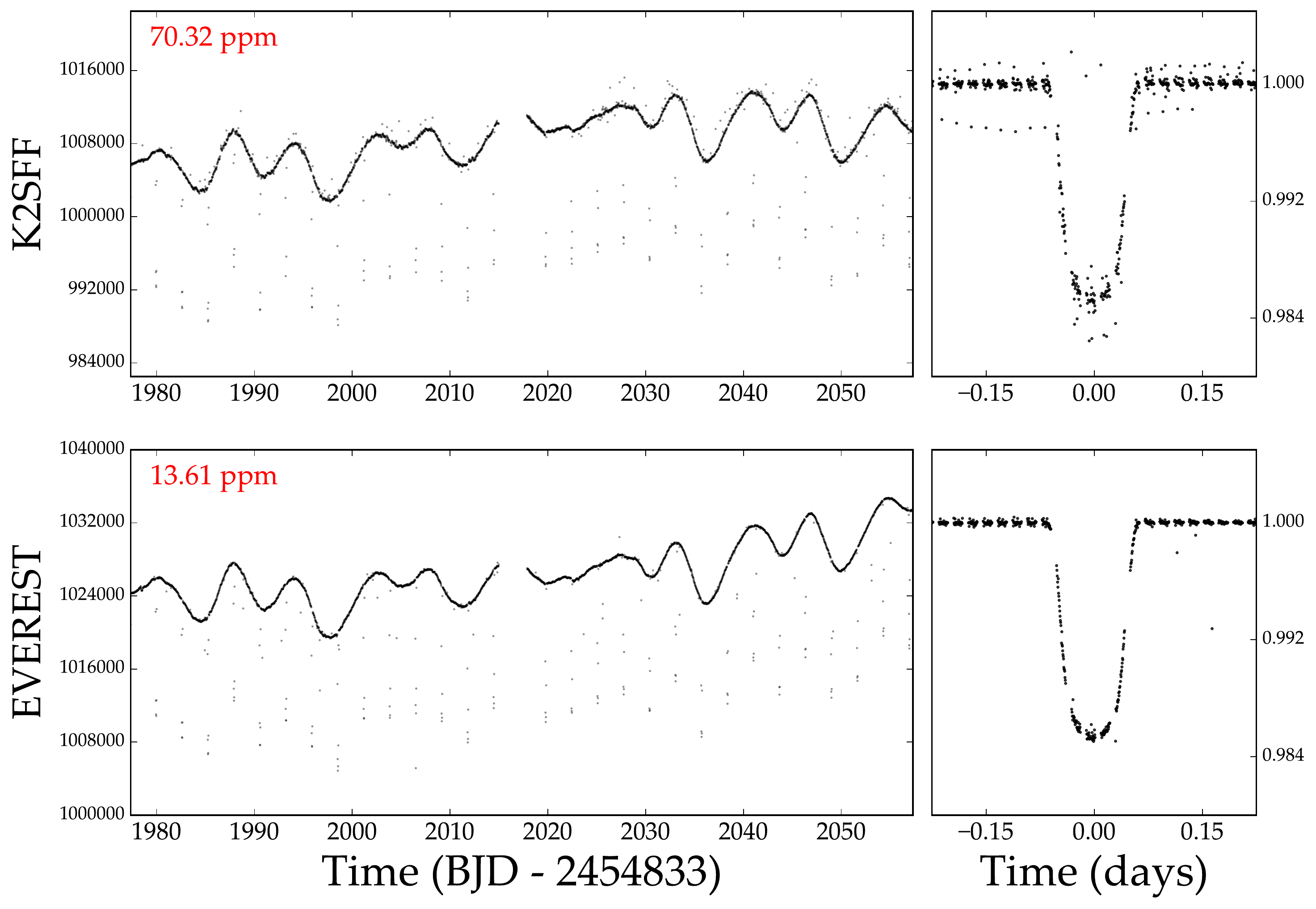}
       \caption{EPIC 201862715, a saturated campaign 1 planet candidate host. As in 
       Figure~\ref{fig:201345483}, we show both the \texttt{K2SFF} and the \texttt{EVEREST 2.0}
       light curves. The CDPP of the \texttt{EVEREST 2.0} light curve is a factor of 5 lower.}
     \label{fig:201862715}
  \end{center}
\end{figure}

In Figures~\ref{fig:201345483} and \ref{fig:201862715} we show two sample light curves
de-trended with both \texttt{K2SFF} and \texttt{EVEREST 2.0}. Figure~\ref{fig:201345483}
shows EPIC 201345483, a faint $\Kp = 15$ planet host.
The CDPP of the \texttt{EVEREST} light curve (bottom) is a factor of 2.4 lower than that of
\texttt{K2SFF} (top), and the light curve has visibly fewer outliers. The folded transit
is shown at the right and is similarly less noisy in the \texttt{EVEREST} light curve.

In contrast, Figure~\ref{fig:201862715} shows a very bright ($\Kp = 10$) saturated 
planet candidate host, EPIC 201862715. With the column-collapsing scheme and the
inclusion of neighboring PLD vectors, \texttt{EVEREST 2.0} is able to achieve a factor
of 5 lower CDPP than \texttt{K2SFF}, as well as considerably fewer outliers.

\subsection{Comparison to Kepler}
\label{sec:kepler}

\begin{figure*}[hbt]
  \begin{center}
      \includegraphics[width=\textwidth]{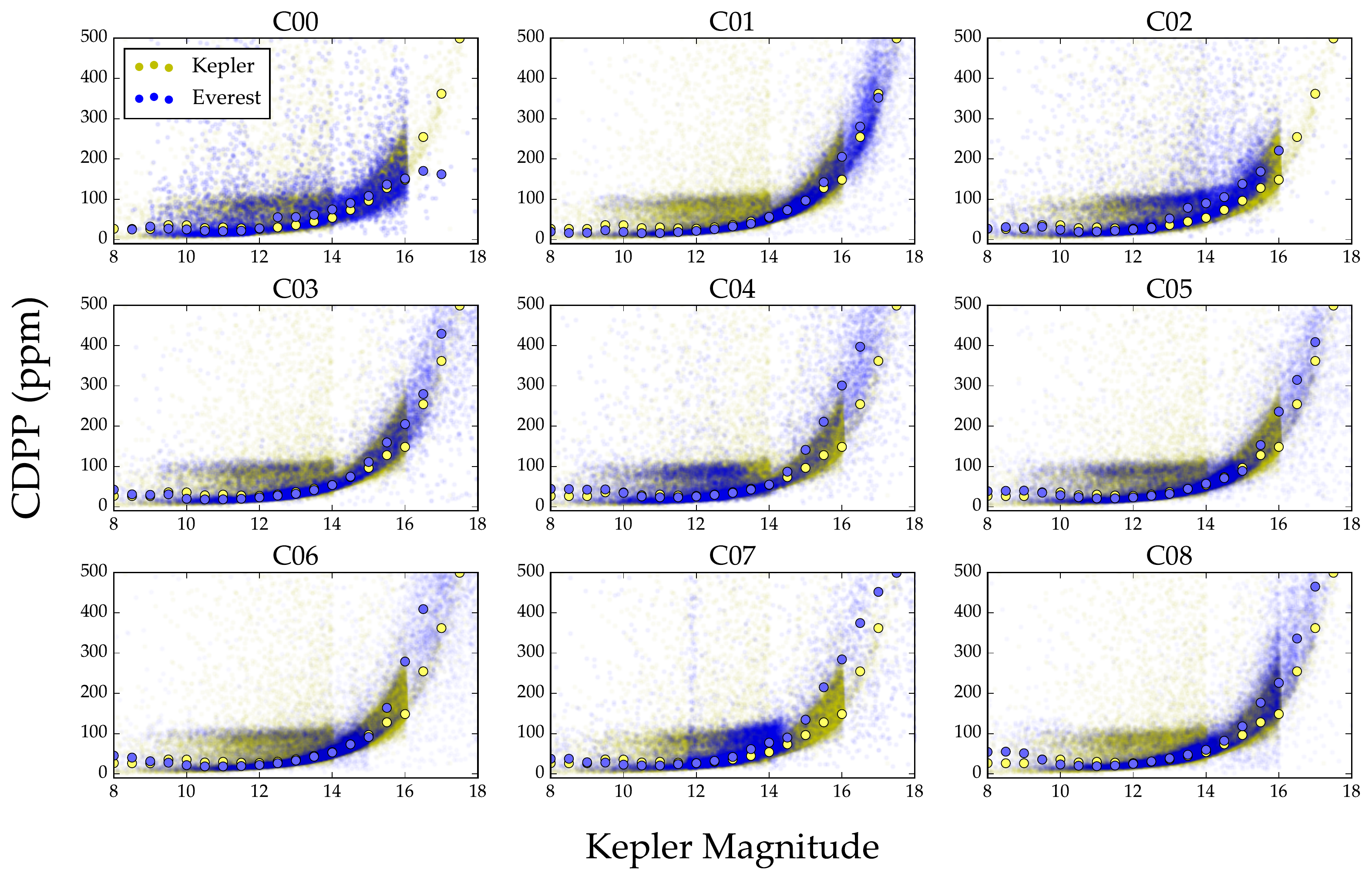}
       \caption{6 hr photometric precision as a function of \emph{Kepler} magnitude $\Kp$ for all 
       stars observed by \emph{Kepler} (yellow dots) and for all $K2$ targets in Campaigns 0-8
       de-trended with \texttt{EVEREST} (blue). The median in 0.5 magnitude-wide bins is indicated
       by yellow circles for \emph{Kepler} and by blue circles for \texttt{EVEREST}. For campaigns
       1, 5, and 6, \texttt{EVEREST} recovers the raw \emph{Kepler} photometric precision down to
       at least $\Kp = 15$; for campaigns 3, 4, and 8, \texttt{EVEREST} recovers the \emph{Kepler} 
       precision down to $\Kp = 14$. Campaigns 0 and 2 have a larger fraction of (variable) giant
       stars, leading to a higher average CDPP, while campaign 7 raw light curves have significantly
       worse precision due to a change in the orientation of the spacecraft and excess jitter.
       }
     \label{fig:cdpp_kepler}
  \end{center}
\end{figure*}

\begin{figure}[hbt]
  \begin{center}
      \includegraphics[width=0.47\textwidth]{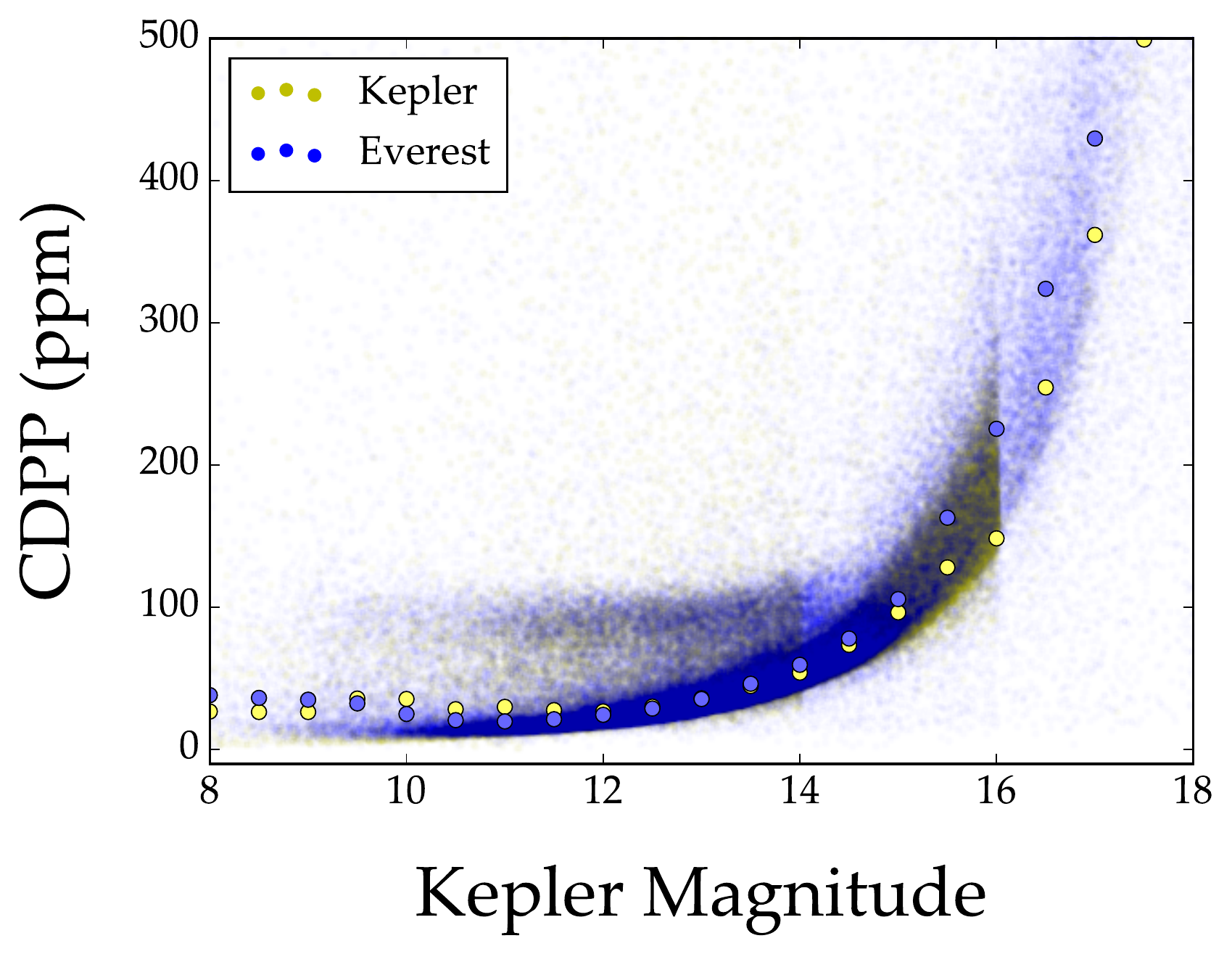}
       \caption{The same as Figure~\ref{fig:cdpp_kepler}, but comparing the CDPP of \emph{all} $K2$ stars
                to that of \emph{Kepler}. \texttt{EVEREST 2.0} recovers the original \emph{Kepler} 
                photometric precision down to at least $\Kp = 14$, and past $\Kp = 15$ for some
                campaigns.}
     \label{fig:cdpp_kepler_all}
  \end{center}
\end{figure}

In Figures~\ref{fig:cdpp_kepler} and \ref{fig:cdpp_kepler_all} we compare the \texttt{EVEREST 2.0}
photometric precision to that of the original \emph{Kepler} mission. Figure~\ref{fig:cdpp_kepler}
shows the CDPP as a function of $\Kp$ for each of the first 9 $K2$ campaigns and 
Figure~\ref{fig:cdpp_kepler_all} shows the comparison for all $K2$ stars. Because of differences
in the raw photometric precision and in the stellar populations across the campaigns, the
results are variable, but for all campaigns except 0, 2 and 7, we recover the original \emph{Kepler}
precision down to at least $\Kp = 14$. For campaigns 1, 5, and 6, we recover the \emph{Kepler}
precision down to at least $\Kp = 15$. This also applies to saturated stars, though in some
campaigns the \texttt{EVEREST 2.0} CDPP is slightly higher for these stars. For stars
dimmer than $\Kp = 15$, the \texttt{EVEREST 2.0} CDPP is within a few tens of percent ---
or less --- than that of \emph{Kepler}.

\section{Additional Remarks}
\label{sec:remarks}

\subsection{Variable Stars}
\label{sec:variables}
In Paper I we discussed how \texttt{EVEREST} often fails to properly de-trend
extremely variable stars, such as RR Lyrae variables and very short period eclipsing
binaries, causing overfitting in many of these light curves. We attributed this to 
there being too much power in the GP model, which
captured both the astrophysical and the instrumental variability, resulting in an
improperly optimized PLD model. However, after considerable experimentation, we
found that this behavior stemmed in large part from our cross-validation scheme.
In Paper I, our cross-validation sets were 13 cadences (6.5 hours) long, and we sought to
minimize the median (proxy) CDPP of all such sets. We chose this timescale
because it is roughly the duration of a typical transit, and in practice
it worked well to minimize overfitting of transits. However, the CDPP as defined in
\S\ref{sec:results} --- the normalized standard deviation in 13 cadence segments
after the application of a high-pass filter --- is not an adequate metric of the
photometric precision for stars that are intrinsically variable on similarly short 
timescales. In other words, RR Lyrae and other extreme variables are dominated by astrophysical
variability on the timescale at which the CDPP is computed, and therefore 
minimizing the CDPP is a recipe for overfitting.

In \texttt{EVEREST 2.0}, we modified our cross-validation scheme (\S\ref{sec:impl_crossval})
in two important ways: we increased the average size of the validation sets to
${\sim}500$ cadences, and we minimized the validation scatter after the subtraction of
a properly optimized GP model. This helps to ensure that we minimize \emph{only} the instrumental
component of the noise. In order to assess the performance of this new scheme, we
visually inspected the de-trended light curves of 100 RR Lyrae stars in campaigns 0--4,
chosen as the targets with the highest probability of being RRab stars
according to the \texttt{K2VARCAT} catalog \citep{Armstrong16}.
Among the \texttt{EVEREST 1.0} light curves, 92/100 had visibly damped
oscillation amplitudes or clearly overfitted stellar variability features. In contrast,
only 44/100 \texttt{EVEREST 2.0} RR Lyrae stars showed any signs of overfitting. For 8
of these, the overfitting occurred only with the inclusion of the second or third order PLD
models. While there are still issues with how the pipeline handles extremely variable
stars, the improvement over version \texttt{1.0} is substantial. 

For comparison, we visually inspected the same stars in the \texttt{K2SFF} and \texttt{K2SC} 
catalogs. 96/100 \texttt{K2SFF} light curves of RRab stars showed signs of overfitting
(dampened oscillation amplitudes, distorted astrophysical signal, 
or significant de-trending artefacts), while only 12/65 (18\%) of \texttt{K2SC} light curves
appeared to be incorrectly de-trended (35 of the 100 stars were in campaigns 0--2 and are
not present in the \texttt{K2SC} catalog). \texttt{K2SC} likely overperforms the other
pipelines for these stars because of its robust GP optimization scheme. Since we optimize
the GP and de-trend the light curve in separate steps, the covariance matrix we use is often
an improper approximation to the true covariance of the astrophysical signal. Progressive
optimization of the GP (\S\ref{sec:impl_gp}) helps to maximize the amount of astrophysical
information captured by the GP model, but a better procedure would be to simultaneously
fit for the GP hyperparameters and the PLD coefficients. This would ensure that the PLD
model captures only instrumental signals and the GP model captures only astrophysical
signals. However, such a method is computationally intractable, since the problem would
no longer be linear. We therefore settle for our linear method, which works extremely
well for stars that do not exhibit extreme variability and has a ${\sim}50\%$ success
rate for those that do. We encourage readers interested in RR Lyrae and other
extreme variables to inspect the light curves of the different pipelines on a target-by-target
basis.

\subsection{Crowded Apertures}
\label{sec:crowded}

\begin{figure}[hbt]
  \begin{center}
      \includegraphics[width=0.47\textwidth]{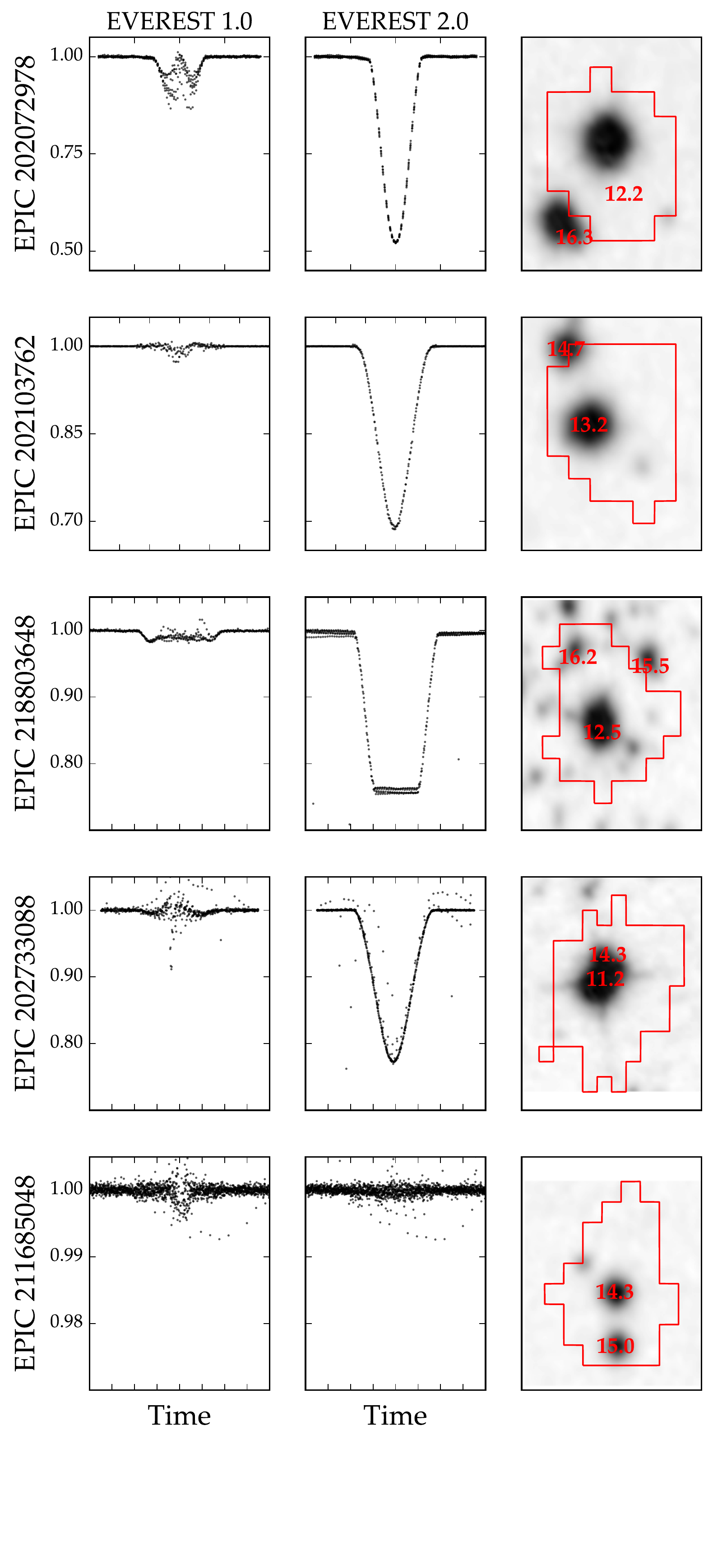}
      \vspace*{-0.5in}
       \caption{Five eclipsing binaries with significant contamination by bright nearby stars. The first 
       two panels in each row show the folded \texttt{EVEREST 1.0} and \texttt{EVEREST 2.0} light curves, respectively,
       and the third shows the POSS high resolution image of
       the target postage stamp with our adopted aperture indicated in red. The magnitudes of the target
       and its bright neighbors are also indicated. While \texttt{EVEREST 1.0} severely overfits the eclipses of
       all five targets, \texttt{EVEREST 2.0} preserves the eclipse depths in all but the last one.}
     \label{fig:crowding}
  \end{center}
\end{figure}

In Paper I we also showed how PLD is not suited to de-trend stars in excessively crowded fields,
since the algorithm will often use pixels from contaminant sources to fit out the target star's astrophysical 
signals. Directly addressing this issue is far less straightforward than mitigating the effects of saturation (\S\ref{sec:impl_saturated})
or extreme astrophysical variability (\S\ref{sec:variables}), as it likely requires accurate modeling and
subtraction of the PSFs of contaminant sources. However, in practice we find that our new de-trending
algorithm is far more robust to overfitting when contaminant sources are present. This is in part
due to the inclusion of PLD regressors from neighboring stars, which reduce the contribution from
potentially contaminated pixels in the target aperture. Our new cross-validation scheme (\S\ref{sec:impl_crossval})
also helps guard against the effects of crowding, since it is better at preventing localized 
overfitting. Because of the large amplitude of the $K2$ spacecraft drift, the amount of contamination
in the target aperture often varies considerably over the duration of the campaign, as nearby sources
move in and out of the aperture. In \texttt{EVEREST 1.0} light curves, this resulted in a time-variable
de-trending power --- and time-variable overfitting --- for crowded targets. As we discussed in \S\ref{sec:impl_crossval},
our new cross-validation method strongly disfavors this behavior.

In Figure~\ref{fig:crowding} we show the light curves of five eclipsing binaries whose crowded apertures
resulted in severe overfitting in the \texttt{EVEREST 1.0} catalog. Shown are the light curves (flux versus time) folded
on the period of the binary and centered on the primary eclipse for \texttt{EVEREST 1.0} (left) and \texttt{EVEREST 2.0} (center).
The right panel shows the Palomar Optical Sky Survey\footnote{\url{http://stdatu.stsci.edu/cgi-bin/dss_form}} (POSS II)
red filter image of the target postage stamp, where contaminant sources are clearly visible in each of the apertures (indicated
by the red contours).

The first three targets (EPIC 202072978, EPIC 202103762, and EPIC 218803648) have relatively bright
contaminant sources centered on or near the edge of the aperture. In Paper I we explained how this
was the ``worst case'' scenario for crowding, as it results in the largest spatial variation of 
astrophysical information from the target across the aperture; as expected, \texttt{EVEREST 1.0}
almost completely fits out their eclipses. \texttt{EVEREST 2.0}, on the other hand, achieves high
de-trending power while preserving the eclipse shapes and depths seen in the raw light curves.
The fourth target (EPIC 202733088) has a bright nearby source inside its aperture, separated by 
less than a pixel. We discussed in Paper I how targets with co-located contaminants are typically
unaffected by PLD overfitting, since the ratio of the two PSFs is roughly constant everywhere.
However, in this case the contaminant is sufficiently detached to result in severe overfitting
in the \texttt{EVEREST 1.0} light curve. As before, \texttt{EVEREST 2.0} correctly de-trends the light curve with
no overfitting. The final target (EPIC 211685048) is an eclipsing binary with deep transits that are completely
fit out in both versions of the pipeline. In this case, two roughly equal magnitude stars are
fully contained in the aperture and separated by ${\sim}3$ pixels, causing PLD to fail despite
the modifications to the algorithm.

As in Paper I, we urge caution when using the \texttt{EVEREST} light curves of crowded targets,
but note that in most of the cases known to us in which the previous version overfitted transits 
and eclipses, \texttt{EVEREST 2.0} succeeds in removing the instrumental noise without overfitting
astrophysical information. Future updates to the pipeline will include a more careful aperture
selection, which can mitigate the effects of crowding for targets like EPIC 211685048.

\subsection{Aperture Losses}
\label{sec:losses}
As we have demonstrated, column-collapsed PLD works extremely well for saturated stars,
provided \emph{all} of the target flux in the saturated columns is used. With this
in mind, we used larger apertures for these stars and
extended them by two pixels at the top and bottom of all saturated columns 
(\S\ref{sec:impl_saturated}). However, for some extremely bright ($\Kp \lesssim 8.5$) 
targets, bleed trails in the saturated columns can extend past the edge of the
target postage stamp. Since most of the astrophysical information in saturated columns
is contained in the two pixels at the top and bottom, this can lead to substantial
information loss. In this case, the column collapsing procedure fails to satisfy the
PLD constraint that all pixels should have the same fractional astrophysical signal
strength, and overfitting can occur. This is the case for EPIC 210703831, a campaign 4
$\Kp = 8.1$ star, whose de-trended light curve displays several discontinuities. These
occur because spacecraft drift results in aperture losses only during parts of the
campaign. We therefore encourage those using the \texttt{EVEREST} catalog to inspect
the postage stamps of extremely bright stars to ensure that aperture losses are not
present.
%
%EPIC 220651426: different de-trending power in each segment.

\section{Using EVEREST}
\label{sec:using}
All \texttt{EVEREST 2.0} $K2$ light curves are available for download as \texttt{FITS} files
from MAST.\footnote{\url{https://archive.stsci.edu/prepds/everest}} As with the 
previous version, we urge users to interface with the catalog via the Python code,
which can be installed following the instructions at 
\url{https://github.com/rodluger/everest}. A detailed description of the data products
and how to use the code is available on the \texttt{github} page. Below, we provide
a brief outline of the \texttt{EVEREST} resources available online.

\subsection{\texttt{FITS} Files}
\label{sec:fits}
Each \texttt{FITS} file contains six extensions. The primary (index \texttt{0}) 
extension consists of a header with miscellaneous target information copied from
the $K2$ target pixel file (TPF). The second (index \texttt{1}) extension contains
a header and a binary table. The header stores miscellaneous
information about the target and the settings used in the de-trending, such as the 
GP hyperparameters and information on the neighboring targets used in the regression.
The binary table stores arrays corresponding to the cadence number (\texttt{CADN}),
the timestamp (\texttt{TIME}), the raw SAP flux (\texttt{FRAW}), the raw SAP flux errors
(\texttt{FRAW\_ERR}), the PLD-de-trended flux (\texttt{FLUX}), the five CBV regressors
(\texttt{CBV01} -- \texttt{CBV05}), and the de-trended flux with the CBV
correction (\texttt{FCOR}). This extension also includes the
original $K2$ \texttt{QUALITY} bit array, with four additional bits that signal
cadences that were masked when computing the model: 
\begin{center}
\begin{tabular}{ |l|l| }
  \hline
  \textbf{23} & Data point is flagged in the $K2$ TPF \\
  \hline
  \textbf{24} & Data point is \texttt{NaN} in the $K2$ TPF \\
  \hline
  \textbf{25} & Data point is an outlier \\
  \hline
  \textbf{26} & \emph{Not used} \\
  \hline
  \textbf{27} & Data point is in a transit (short cad. only) \\
  \hline
\end{tabular}
\end{center}
When using the \texttt{EVEREST} light curves for science purposes, it is important
to properly reject these outliers. Data flagged with bit \texttt{23} is typically 
affected by cosmic rays or detector anomalies, and can usually be ignored. In
some cases, however, deep transits or eclipses can be mistaken for detector
anomalies and are incorrectly flagged in the original TPF, so a visual inspection
of the light curves is recommended. Data flagged with bit \texttt{24}, on the other
hand, is missing in the original TPF and can thus be safely ignored. Finally, bit
\texttt{25} usually corresponds to instrumental outliers that were not properly
de-trended with PLD. However, transits, flares, and other short timescale astrophysical
features will also be flagged with this bit, so this data should \emph{not} be
blindly excluded when performing transit searches.
  
The four remaining \texttt{FITS} extensions include the PLD regressors, the target aperture, 
and additional information used internally by the Python code.

\subsection{Data Validation Summaries}
\label{sec:dvs}
Each target in the catalog also has an associated data validation summary (DVS), a
\texttt{PDF} document showing the raw, de-trended, and CBV-corrected light curves, 
as well as cross-validation diagnostic plots such as those in Figure~\ref{fig:crossval}
and a high resolution POSS image of the target aperture like those shown in 
Figure~\ref{fig:crowding}.

\subsection{Python Code}
\label{sec:python}
The \texttt{EVEREST} code can be installed using the package managing system \texttt{pip}:
\begin{lstlisting}[language=bash]
pip install everest-pipeline
\end{lstlisting}
or directly from source by following the instructions on the \texttt{github} page. 
The primary way of interfacing
with the catalog is to instantiate an \texttt{Everest} object:
\lstset{emph={EPIC}, emphstyle=\itshape}
\begin{lstlisting}[language=Python]
import everest
star = everest.Everest(EPIC)
\end{lstlisting}
where \texttt{EPIC} is the EPIC number of the target. These lines download the
target's \texttt{FITS} file and populate
the \texttt{star} object with the de-trending information and light curve arrays
(\texttt{time}, \texttt{flux}, \texttt{fcor}, etc.).
The \texttt{QUALITY} flags \texttt{23}, \texttt{24}, \texttt{25}, and \texttt{27} are 
used to generate the \texttt{badmask}, \texttt{nanmask}, \texttt{outmask}, and 
\texttt{transitmask} arrays, respectively; these arrays contain the indices of all
data points whose corresponding bit is flagged.

As we discussed in \S\ref{sec:inj}, it is important that transits are properly masked
when computing the PLD model; otherwise, slight overfitting may occur. Users should 
therefore always mask known transits and re-compute the PLD model, as follows:
\lstset{emph={time, period, duration}, emphstyle=\itshape}
\begin{lstlisting}[language=Python]
star.mask_planet(time, period, duration)
star.compute()
\end{lstlisting}
where \texttt{time} is the time of first transit (in units of $\mathrm{BJD} - 2454833$),
\texttt{period} is the transit period (in days), and \texttt{duration} is the full
transit duration (also in days). Re-computing the model takes no more than 
a few seconds for long cadence light curves.

Users can also easily change the number of CBVs used to correct the light curve:
\lstset{emph={n}, emphstyle=\itshape}
\begin{lstlisting}[language=Python]
star.cbv_num = n
star.compute()
\end{lstlisting}
where \texttt{n} is the desired number of CBVs (0 -- 5).

The code also implements various visualization routines, which are described in
the documentation linked on the \texttt{github} page.

\section{Conclusions}
\label{sec:conclusions}
We have presented \texttt{EVEREST 2.0}, an update to the \texttt{EVEREST} pipeline
\citep{Luger16} for removing instrumental noise from $K2$ photometry. In version
\texttt{1.0}, we constructed a linear model from the principal components of
products of the fractional pixel fluxes in the aperture of each star, a variant
of a method known as pixel level decorrelation \citep[PLD,][]{Deming15}. Here,
we regress on \emph{all} PLD vectors, imposing Gaussian priors on the model
weights to prevent overfitting. We additionally include the PLD vectors of bright
neighboring stars to increase the signal-to-noise ratio of the regressors and
enhance the predictive power of the model. We developed a fast gaussian process (GP)
regression scheme to de-trend all stars in the $K2$ catalog, achieving lower
combined differential photometric precision (CDPP) than in version \texttt{1.0},
by ${\sim}10\%$ for bright stars and ${\sim}20\%$ for faint stars. We also
adapted PLD to work for saturated stars, yielding comparable de-trending power,
and stars observed in short cadence mode, yielding higher photometric precision
on 6 hr timescales than their long cadence counterparts.
We further find that the inclusion of neighboring PLD vectors and a
more conservative cross-validation scheme enhance the pipeline's robustness
to overfitting, particularly for highly variable stars.

\texttt{EVEREST 2.0} light curves have higher photometric precision than the two
other publicly available catalogs, \texttt{K2SFF} \citep{VanderburgJohnson14}
and \texttt{K2SC} \citep{Aigrain16}, at all $\Kp$ magnitudes. For faint stars,
\texttt{EVEREST 2.0} has ${\sim}40\%$ lower CDPP than \texttt{K2SFF} and
${\sim}25\%$ lower CDPP than \texttt{K2SC}; for bright unsaturated stars, the CDPP
improvement is ${\sim}20\%$ compared to both pipelines. For saturated stars,
\texttt{EVEREST} outperforms both pipelines, but by a smaller margin. We also
find that \texttt{EVEREST} light curves have, on average, 100--300 fewer outliers
than those of other pipelines, owing primarily to the ability of PLD to correct
data collected during thruster firing events.

When compared to the original \emph{Kepler} mission, \texttt{EVEREST 2.0}
recovers \emph{Kepler} photometry on average to $\Kp \approx 14.5$, and past
$\Kp = 15$ for some campaigns. For dimmer stars, the CDPP is within a few tens
of percent of that of \emph{Kepler}. \texttt{EVEREST} light curves should thus enable 
continued high-precision transiting exoplanet and stellar variability science 
for the vast majority of $K2$ stars as if they had been observed by the original
four-wheeled mission.

The \texttt{EVEREST 2.0} catalog of de-trended light curves is publicly available
at \url{https://archive.stsci.edu/prepds/everest}.
As with the previous version of the code, \texttt{EVEREST 2.0} is open source under
the MIT license and available at \url{https://github.com/rodluger/everest}.
%
% -*-*- TODO -*-*-
%
%, with a static release of the code used to generate the catalog archived at 
%\url{http://dx.doi.org/XX.XXX/XXXXX}. 
%
%
The reader is encouraged to use this code
to interface with the \texttt{EVEREST} catalog and to customize the de-trending
of targets of interest, particularly for masking transits to remove biases in
the depth due to overfitting. We have implemented each of the PLD models discussed
above (\texttt{rPLD}, \texttt{nPLD}, \texttt{pPLD}) in a general framework that
can be adapted to different missions, including \emph{Kepler} and the upcoming
\emph{TESS}. For more information, refer to the documentation linked 
on the \texttt{github} page.

\acknowledgments{
We would like to thank Benjamin Pope and Tsevi Mazeh
for their useful comments and suggestions.
R.L., R.B., and E.A. acknowledge support from NASA grant
NNX14AK26G and from the NASA Astrobiology Institute's
Virtual Planetary Laboratory Lead Team, funded through
the NASA Astrobiology Institute under solicitation
NNH12ZDA002C and Cooperative Agreement Number
NNA13AA93A. E.A. acknowledges additional support from
NASA grants NNX13AF20G and NNX13AF62G. E.K.
acknowledges support from an NSF Graduate Student Research
Fellowship. This research used the advanced computational,
storage, and networking infrastructure provided by the Hyak
supercomputer system at the University of Washington.
}

\bibliographystyle{apj}
\bibliography{everest}
\end{document}